\documentclass[a4paper,11pt]{article}
\pdfoutput=1 % if your are submitting a pdflatex (i.e.~if you have
\usepackage{jheppub,amsthm} % for details on the use of the package, please
\usepackage[utf8]{inputenc}

\usepackage{booktabs,multirow, quotes}
\usepackage{chngcntr}
\usepackage[english]{babel}
\usepackage{amsmath,amssymb,graphicx,xcolor,alltt,tikz,subcaption}
\usepackage{framed}

\theoremstyle{remark}

\theoremstyle{theorem}

\makeatletter
\def\@fpheader{\ }
\makeatother

\counterwithout{figure}{section}
\title{Irrational CFTs from coupled anyon chains with non-invertible symmetries?}

\author{Ant\'onio Antunes$^1$, Junchen Rong$^\tau$}

\affiliation{$^1$Laboratoire de Physique de l'\'Ecole Normale Sup\'erieure, Universit\'e  PSL, CNRS, Sorbonne Universit\'e, Universit\'e  Paris Cit\'e, 24 rue Lhomond, F-75005 Paris, France\,, and\\
   Centro de F\'isica do Porto e Departamento de F\'isica e Astronomia, Faculdade de Ci\^encias da Universidade
 do Porto, Rua do Campo Alegre 687, 4169-007, Porto, Portugal\\
$^\tau$CPHT, CNRS, \'Ecole Polytechnique, Institut Polytechnique de Paris, Palaiseau, France}

\emailAdd{antonio.antunes@ens.fr, junchen.rong@polytechnique.edu}

\abstract{Irrational CFTs in 1+1d with a discrete spectrum and no conserved currents other than the stress-tensor are expected to be generic, unsolvable by standard methods, and hard to construct explicitly. We introduce a lattice model that realizes a candidate for such a CFT as a conformal phase of matter without fine-tuning. The model is constructed by coupling $N\geq3$ golden anyon chains together, preserving $N$ copies of the Fibonacci non-invertible symmetry. We use the MPS/DMRG approach to study this model numerically, which allows us to calculate the corresponding conformal data, obtaining hints of its irrationality. Along the way, we characterize the phase diagram for $N=2$ coupled chains where we identify a weakly first-order phase transition as well as critical points that we are able to identify with known rational CFTs, except for one case. We also provide an extensive list of rational CFTs with $1<c<2.1$. }

\begin{document} 
	
	\maketitle
	
	\flushbottom

\flushbottom

\section{Introduction}
\label{sec:intro}
Rational two-dimensional conformal field theories (RCFTs) are beacons of exact solvability in a sea of intractable models. These theories, which have a finite number of chiral algebra primaries and rational scaling dimensions $\Delta$ and central charges $c$ \cite{Vafa:1988ag}, describe, among many other things, continuous phase transitions of 2d statistical systems and of 1+1d quantum many-body systems.\footnote{
Such quantum critical points are generally described by a broad class of Lifschitz critical phases with a dynamical critical exponent $z$ not necessarily equal to 1. 
The particularly nice subclass of our interest has $z=1$ and therefore emergent relativistic conformal symmetry.} 
While a vast body of work starting with \cite{Belavin:1984vu} describes these exactly solvable RCFTs, much less is known about irrational CFTs with a discrete infinite list of primary scaling dimensions, especially when they do not lie in continuous families and only have the stress-tensor as a conserved current.\footnote{Conformal manifolds, which are only known to occur in the presence of spin-1 or spin-3/2 currents, often have irrational points, but these are not of the type we want to study in this paper. See however \cite{Cordova:2023qei} for a discussion emphasizing non-invertible symmetries in this context.} A special class of these CFTs, those that have a vanishing twist gap, i.e. where excluding the Virasoro vacuum module one has $\textrm{inf} \, (\Delta-J)=\tau_{\textrm{gap}}>0$ are particularly interesting. They are expected to be generic, to satisfy general bootstrap constraints \cite{Hellerman:2009bu,Collier:2016cls,Kusuki:2018wpa,Benjamin:2019stq,Collier:2019weq,Pal:2022vqc,Pal:2023cgk,Dey:2024nje,Pal:2025yvz} and include holographic CFTs dual to quantum gravity in AdS$_3$ \cite{Maloney:2007ud,Hartman:2014oaa}.

In spite of these general results, there are essentially no rigorous constructions of such theories. Fortunately, this is not because of some fundamental theoretical obstruction. Starting with the work of \cite{Dotsenko:1998gyp} and as recently revisited in \cite{Antunes:2022vtb,Antunes:2024mfb}, it is reasonably straightforward to construct candidates for such theories by coupling multiple copies of solvable CFTs, like Virasoro minimal models, by judiciously chosen weakly relevant operators and flowing to non-trivial infrared fixed points which can be described with conformal perturbation theory \cite{Zamolodchikov:1987ti}. Non-perturbative studies of such CFTs can take (at least) two natural paths:
\begin{itemize}
    \item Finding continuous phase transitions in 2d statistical lattice models. This was done in \cite{Dotsenko:1998gyp} where three coupled classical 3-state Potts models were studied using Monte-Carlo and transfer-matrix techiques.
    \item Finding continuous phase transitions in 1+1d quantum many-body systems, which can be studied using modern DMRG techniques with an MPS variational ansatz. This is the approach we will take in this paper.
\end{itemize}
Both these approaches have a natural hurdle to their application: The fixed points above are often multi-critical, i.e. they require tuning multiple relevant parameters. This limitation was overcome in \cite{Dotsenko:1998gyp} by imposing additional lattice duality symmetries at the level of the Hamiltonian. In the modern language, this amounts to \textit{preserving additional non-invertible symmetries at the microscopic level}, and we will make extensive use of these non-invertible symmetries in our 1+1d quantum Hamiltonian \cite{Feiguin:2006ydp,Shao:2023gho,Seiberg:2023cdc}.

We will study a lattice model that preserves the symmetry  
\begin{equation}
   \mathcal{C}=(\rm{Fibonacci})^N \rtimes S_N\,,
\end{equation}
which is both \textit{non-abelian and non-invertible}.\footnote{Another example of such a structure that has appeared before in the literature is the Haagerup fusion category \cite{Huang:2021nvb,Corcoran:2024eeh,Bottini:2025hri,Hung:2025gcp}.}
Here, the Fibonacci symmetries are associated to topological defect lines (TDLs) $W^{(a)}$ satisfying the fusion rule
\begin{equation}
\label{Fibonaccifusion}
    W^{(a)}\times W^{(a)}=1+W^{(a)}\,,
\end{equation}
where $1$ denotes the identity TDL.\footnote{See \cite{Copetti:2024rqj,Nakayama:2024msv,Ambrosino:2025yug} for the use of non-invertible symmetries in describing RG flows starting from a single (non-unitary) minimal model.} 
To realize this on the lattice, we will write a local Hamiltonian
\begin{equation}
\label{latticeNHamilt}
    \hat{H}_{N-\textrm{Fib}} = J\sum_{i=1}^{L}\sum_{a=1}^N \hat{H}_i^{(a)} + K \sum_{i=1}^{L}\sum_{a\neq b}^N \hat{H}_i^{(a)}\hat{H}_i^{(b)}\,,
\end{equation}
where each local contribution to the $L-$site Hamiltonian ($i=1,\dots,L$) commutes with a non-invertible symmetry generator satisfying the same algebra as the TDLs above.\footnote{An exact microscopic symmetry constrains the IR CFT provided it is not
spontaneously broken. For the Fibonacci symmetry this gives a sharp diagnostic:
a broken phase cannot have a one-dimensional vacuum sector, since a
one-dimensional module over \(W^2=1+W\) would require a non-negative integer
solution of \(n^2=1+n\). Therefore spontaneous breaking would produce additional
vacua, visible in the finite-size spectrum as states with \(\Delta\to0\). We do
not observe such states at the candidate critical points.} For each copy, this Hamiltonian
\begin{equation}
    \hat{H}_{\textrm{Fib}} = J\sum_{i=1}^{L} \hat{H}_i\,,
\end{equation}
acts on the Hilbert space of the so-called golden anyon chain~\cite{Feiguin:2006ydp}, where one assigns a qubit at each lattice site but imposes adjacency constraints compatible with the Fibonacci fusion rules \eqref{Fibonaccifusion}, making this a non-factorizable Hilbert space. 
Remarkably, this model flows precisely to the tricritical Ising CFT for any value of $J>0$, and flows to the three-state Potts model CFT for $J<0$.
Adding the $K$-term allows us to couple $N$ copies of these CFTs together and study the corresponding RG flows.

When $J>0$, in field theory language, we will study the following RG flow discussed in \cite{Chang:2018iay}:
\begin{align}\label{actionfield}
    S=\sum_{a=1}^{N} S_{\rm tri-Ising}^{(a)} + K \int d^2x \sum_{a\neq b}\sigma'^{(a)}\sigma'^{(b)}+{\rm irrelevant}\,,
\end{align}
where we couple $N$ copies of the tricritical Ising CFT ($a=1,\dots,N$) with $c=7/10$ via a relevant two-copy interaction of dimension $\Delta_K=7/4$, obtained by multiplying two relevant $\mathbb{Z}_2$ odd operators $\sigma'^{(a)}\equiv\phi_{(2,1)}^{(a)}$, where the subscript $(2,1)$ denotes the usual Kac labels. 
In fact, the full fusion category symmetry of this model is given by
\begin{equation}
   ( (\rm{Fibonacci})^N \rtimes S_N)\times \mathbb{Z}_2\,,
\end{equation}
which includes a diagonal $\mathbb{Z}_2$ symmetry.
It was conjectured in \cite{Chang:2018iay} that the above theory flows to an irrational CFT for $N\geq3$.
Instead, when $J<0$, we can formally write the corresponding field theory as 
\begin{align}\label{actionfield2}
    S=\sum_{a=1}^{N} S_{\rm Potts-3}^{(a)} + K \int d^2x \sum_{a\neq b} Z^{*(a)}Z^{(b)} +{\rm other~irrelevant~terms}\,,
\end{align}
where the $Z$ operator is charged under the $\mathbb{Z}_3$ symmetry of the Potts model. 
It is also invariant under the Fibonacci symmetry. The $K$-term deformation has scaling dimension $\Delta_K=8/3$ in this case, which is irrelevant. 
For small $K$, this should not change the IR behavior. 
We will give evidence that the theory \eqref{actionfield2} is UV completed by a CFT with
\begin{equation}
   ( (\rm{Fibonacci})^N \rtimes S_N)\times S_3\,
\end{equation}
symmetry by studying the lattice model, at least in the case $N=2$.  Let us emphasize the role of the anyon-chain realization in the study of these CFTs. The same infrared fixed points may well be reachable from other coupled lattice regularizations of the nearby critical theories. The advantage of the anyon-chain realization is that the Fibonacci non-invertible symmetries are exact microscopic symmetries. In a generic coupled lattice model preserving only the ordinary invertible symmetries, additional relevant perturbations
allowed by those symmetries would mix in and would have to be tuned away. Thus the anyon-chain construction gives a symmetry-protected route to the fixed point rather than merely a different lattice discretization of the same continuum perturbation.

We will study the lattice Hamiltonian \eqref{latticeNHamilt} for $N=2,3$ using standard DMRG methods based on the variational MPS ansatz, determining two main types of observables:
\begin{itemize}
    \item \textit{Scaling dimensions of local operators} These can be obtained through the state-operator correspondence, which relates them to energies of the anyon chain on the circle. The key equation is~\cite{JLCardy_1984,PhysRevLett.56.742,CARDY1986186}
    \begin{equation}\label{energyscaling}
         E=E_{\infty}L+\frac{2\pi v}{L}\left(\Delta-\frac{c}{12}\right)+\ldots\,,
    \end{equation}
    where $E_{\infty}$ and $v$ are non-universal constants. The DMRG procedure finds the ground state and its energy and by subsequent orthogonalization gives access to excited states.
    \item \textit{Central charge} While it can in principle be obtained from finite-size scaling of the energy spectrum, it is more easily extracted from the entanglement entropy between two sub-regions of length $\ell$ and $L-\ell$ in the ground state of the system, which is directly provided by the DMRG. The entanglement entropy at the critical point follows the law
    \begin{equation}\label{centralchargeL}
        S_{\textrm{Entanglement}}= \frac{c}{3} \log[l_{\rm conformal}]+ s_1,\quad {\rm with}\quad l_{\rm conformal}=\frac{L}{\pi} \sin(\frac{l}{L}\pi)\,,
    \end{equation}
    with $s_1$ a non-universal constant. 
\end{itemize}
By analyzing the observables above, we arrive at our main results:
\begin{itemize}
    \item For $N=2$ chains, there is a weakly first-order phase transition when $J>0$ and $K<0$. Instead, for $K
    >0$, the theory is gapped.
    Extending the analysis to the $J<0$ region, we find a conformal phase corresponding to two decoupled 3-state Potts models, as well as two special points corresponding to non-trivial CFTs with $c\approx1.35$ and $c\approx1.77$. The first theory corresponds to the well-known coset model $(SU(2)_3 \times SU(2)_3)/SU(2)_6$, but we were not able to identify the second theory. Whether this critical point is rational or not remains an open question.
    \item For $N=3$, there is once again a weakly first-order phase transition for $J>0$ and $K<0$, but for $K>0$ we now find a non-trivial conformal phase with $c=2.10\pm0.03$. We conjecture this to be an irrational CFT at the end of a short RG flow described by \eqref{actionfield}. The central charge of this theory can be estimated from conformal perturbation theory to be $c_{\rm{IR}}\approx2.09$ which is well within our error bars.
\end{itemize}

The structure of the rest of the paper is as follows. In Section \ref{sec:Single} we review the single golden anyon chain, discussing the Hilbert space and the Hamiltonian in detail. Making use of the non-invertible symmetry, we identify the $\sigma'$ deformation and study the gapped phase obtained by turning on this operator in the Hamiltonian. In Section \ref{sec:two} we consider the two-coupled chains discovering a weakly first order phase transition. We also probe the full phase diagram, and analyzing the invertible and non-invertible symmetries we discuss two non-trivial CFTs that occur at special values of the couplings (one rational, one unidentified) as well as a conformal phase described by decoupled Potts models. In Section \ref{sec:three} we extend the analysis to the $N=3$ case once again finding a weakly first order phase transition for one sign of the coupling but discovering a non-trivial conformal phase for the other sign. We conjecture this to be an irrational CFT that is reached in the IR of a short RG flow from decoupled tricritical Ising models as we argue from conformal perturbation theory. Some indirect evidence is given for this claim in Section \ref{sec:list}, where we compare the non-trivial fixed point to a list of known RCFTs, finding no obvious candidates for a match. We conclude in Section \ref{sec:Future} were we discuss possible future directions. The DMRG calculations in this paper are performed using the ITensor package~\cite{itensor,itensor-r0.3}.

\medskip
\textbf{Note added:} In an earlier version of this paper the weakly first-order phase transition of the $N=2$ model mentioned above was misidentified as a non-trivial CFT with $c\approx1.15$. A similar misidentification was also made for the $N=3$ model. We thank Hubert Saleur for first bringing this to our attention. See also \cite{Blakeney:2025ext} for a detailed discussion of the issue of large correlation lengths in anyon chains, including the $(\rm{Fibonacci})^2$ case.

\section{Single anyon chain}
\label{sec:Single}

In this section, we will review the so-called golden anyon chain (also known as Fibonacci anyons) first constructed in \cite{Feiguin:2006ydp} and nicely summarized in \cite{Trebst_2008}. This is a model of non-abelian anyons which has been proposed to be relevant in the fractional quantum Hall effect at filling fraction $\nu=12/5$ (the so-called Read-Rezayi state \cite{Read:1998ed}), as well as in topological quantum computing as it is the simplest computationally complete topological system \cite{Freedman:2001eqc}. For our purposes, the most important feature of this model is that it realizes a conformal phase in the $c=7/10$ tricritical Ising universality class and the $c=4/5$ three-state Potts model universality class.
\subsection{Review of the golden anyon model}
\label{ssec:Review}
The Hilbert space of the golden anyons is defined as the set of degenerate ground states in the fusion space of a one-dimensional chain of $L$ Fibonacci anyons. Each Fibonacci anyon can belong to one of two types: The trivial anyon denoted by $1$ and a charged non-trivial anyon denoted by $\tau$. When two such anyons are fused, they obey the following rules
\begin{equation}
    1\times1=1 \,, \quad 1\times \tau=\tau\times1=\tau\,, \quad \tau \times \tau=1+\tau\,,
\end{equation}
where $\times$ denotes the fusion operation, and one immediately notices the non-invertibility of the $\tau$ line.  Indeed, the quantum dimension of this line $d_\tau=\langle \tau \rangle= \varphi= (1+\sqrt{5})/2$ is greater than one, giving the non-abelian character to the anyon model, as well as justifying its `golden' moniker.  Following these rules, when we fuse $L$ of the $\tau$ anyons, a specific state is labeled by a sequence of intermediate $1$'s and $\tau$'s such that a $1$ is always followed by a $\tau$, as is required by the fusion rules. We write such a state as $|x_1,\dots,x_{L-3}\rangle$ with $x_i \in \{1,\tau\}$. Importantly, even writing such a state implicitly assumes a choice on a basis which amounts to picking in which order we fuse the anyons. We will work in the canonical basis were we fuse the anyons sequentially from left to right, and a state $|x_1,\dots,x_{L-3}\rangle$ is depicted schematically in Fig.~\ref{fig:Fibo}.

\begin{figure}[h]
\centering
\begin{tikzpicture}[thick, baseline={(0,-0.5)}, every node/.style={inner sep=0.5pt}]

  % Horizontal line
  \draw (-0.7,0) -- (5.7,0);
  
  % Vertical lines and tau labels
  \foreach \i in {0,...,4} {
    \draw (\i+0.5,0) -- (\i+0.5,1.5);
    \node at (\i+0.5,1.8) {$\tau$};
  }
  
  % Chi labels
  \node at (1.0,0.3) {$x_1$};
  \node at (2.0,0.3) {$x_2$};
  \node at (3.0,0.3) {$\dots$};
  \node at (4.0,0.3) {$x_{L-3}$};
   \node at (-1.0,0) {$\tau$};
      \node at (6.0,0) {$\tau$};
  % Dotted line for omitted chi's
\end{tikzpicture} 
  \caption{Fusion diagram for the Hilbert space of ground states of $L$ Fibonacci anyons in the left-to-right fusion basis.}  
  \label{fig:Fibo}
\end{figure}
  It is straightforward to count the dimension of this Hilbert space. Let $\textrm{dim} \,\mathcal{H}_L$ be the number of ground states. By using the fusion rules, either the first fusion is given by $\tau$ in which case there are $\textrm{dim} \,\mathcal{H}_{L-1}$ states or it is given by $1$ in which case there are $\textrm{dim} \,\mathcal{H}_{L-2}$ states. This leads to the recursion relation
  \begin{equation}
      \textrm{dim} \,\mathcal{H}_{L}=\textrm{dim} \,\mathcal{H}_{L-1}+\textrm{dim} \,\mathcal{H}_{L-2}\,,
  \end{equation}
  which implies that the dimension of the Hilbert space is
  \begin{equation}
      \textrm{dim} \,\mathcal{H}_{L} = \textrm{Fib}_L\,,
  \end{equation}
  which is the $L$th Fibonacci number. In the thermodynamic limit $L\to\infty$, the dimension grows exponentially $    \textrm{dim} \,\mathcal{H}_{L} \sim \varphi^L$ but with a non-integer base. This follows from the fact that the Hilbert space does not factorize due to the fusion constraints. 
  
  \medskip
  
The highly degenerate and gapped ground state of the non-interacting anyons can be modified by adding interactions that lead to a non-trivial collective ground state that we will be able to make gapless through simple dynamics. The idea is to define an `anti-ferromagnetic' Hamiltonian that privileges the fusion of two $\tau$ anyons into the trivial $1$ channel. This is in complete analogy with a Heisenberg interaction that we review here for clarity. Let us consider a periodic chain of $L$ spin$-1/2$ variables with the usual Heisenberg Hamiltonian
  \begin{equation}
      \hat{H}_{\textrm{Heisenberg}} =  J \sum_{i=1}^L (\vec{S}_i \cdot \vec{S}_{i+1})\,.
  \end{equation}
By rewriting the scalar product as a difference of total and individual angular momentum, one finds
\begin{equation}
    \hat{H}_{\textrm{Heisenberg}} =  \frac{J}{2} \sum_{i=1}^L \left(\hat{\Pi}_{i,i+1}^{(0)}-\frac{3}{2}\right)\,,
\end{equation}
where $\Pi_{i,i+1}^{(0)}$ denotes a projector onto the singlet (spin 0) sector in the $\tfrac{1}{2} \otimes \tfrac{1}{2}= 0 \oplus 1$ tensor product of each neighbouring pair of spins. Indeed, for $J>0$ we see that the chain is anti-ferromagnetic as it favours the formation of singlets.

Back to our anyon chain, it is not immediately clear how to favor fusion onto the $1$ channel since our fusion tree does not represent fusion in terms of the adjacent external anyons, but instead fuses an external anyon with an internal one. This however is simply due to our choice of basis, which can easily be locally modified. Let us denote this change of basis matrix by $F$. Diagramatically, this change of basis is represented by the $F-$move of Fig.~\ref{fig:Fmove}.

\begin{figure}[ht]
  \centering
  \begin{tikzpicture}[baseline={(0,-0.5)}, thick, every node/.style={font=\normalsize}]
    % LEFT SIDE: (x_{i-1}, x_i, x_{i+1})
    \draw (-3.5,0) -- (-1.5,0);  % Horizontal base
    \draw (-3.,0) -- (-3.,1.5); % First vertical
    \draw (-2,0) -- (-2,1.5); % Second vertical
    \node at (-3,1.7) {$\tau$};
    \node at (-2,1.7) {$\tau$};
    \node at (-3.5, -0.3) {$x_{i-1}$};
    \node at (-2.5, -0.3) {$x_i$};
    \node at (-1.5, -0.3) {$x_{i+1}$};

    % Equal sign and F-symbol
    \node at (0, 0.5) {$= \sum_{x_i'} \left( F^{x_{i+1}}_{x_{i-1} \, \tau \, \tau} \right)^{x_i'}_{x_i}$};

    % RIGHT SIDE: F-move with trivalent vertex
    \draw (2,0) -- (4,0);  % Horizontal base
    \draw (3,0) -- (3,0.75); % Vertical part of Y
    \draw (3,0.75) -- (2.5,1.5); % Left tau
    \draw (3,0.75) -- (3.5,1.5); % Right tau
    \node at (2.5,1.7) {$\tau$};
    \node at (3.5,1.7) {$\tau$};
    \node at (2.7,0.5) {$x_i'$};
    \node at (2, -0.3) {$x_{i-1}$};
    \node at (4, -0.3) {$x_{i+1}$};
  \end{tikzpicture}
  \caption{F-move changing the fusion basis of four anyons from intermediate channel $x_i$ to $x_i'$.}
  \label{fig:Fmove}
\end{figure}
  The $F-$matrix is uniquely fixed by consistency through the so-called pentagon equation. In practice, the only fusion space that is more than 1-dimensional is the one for four $\tau$ anyons, so the only non-trivial F-matrix is given by
  \begin{equation}
      F_{\tau\tau\tau}^\tau = \begin{pmatrix}
1/\varphi & 1/\sqrt{\varphi} \\
1/\sqrt{\varphi} & -1/\varphi 
\end{pmatrix}\,.
  \end{equation}

Upon changing to the $x'_i$ basis, it is straightforward to project the fusion onto the singlet channel. He can therefore write our Heisenberg-like Hamiltonian as
\begin{equation}
    \hat{H}_{\textrm{Fib}}= J\sum_{i=1}^L \hat{H}_i \equiv J\sum_{i=1}^L F \, \hat{\Pi}_{i,i+1}^{(1)}F^{-1}\,,
\end{equation}
which means we have an effectively 3-site interaction Hamiltonian in the original $x_i$ basis. After some massaging, we can write
\begin{align}
\hat{H}_i= & \left(n_{i-1}+n_{i+1}-1\right) \nonumber\\
& -n_{i-1} n_{i+1}\left(\varphi^{-3 / 2} \sigma_i^x+\varphi^{-3} n_i+1+\varphi^{-2}\right),
\end{align}
where $n_i=\frac{1}{2}(1-\sigma_i^z)$. Crucially, $H_i$ commutes with the Fibonacci topological symmetry, which acts on the lattice by adding an additional $\tau$ anyon parallel to the spine of the fusion diagram and fusing it past each external anyon through successive $F-$moves. In practice, since we want to simulate this model with DMRG, we need to be able to work with a factorized Hilbert space (with $\{1,\tau\}$ playing the role of a spin-1/2 d.o.f) and impose the fusion constraints dynamically. This is easily achieved by adding strong nearest neighbour repulsive interaction between the $1$ particles, 
\begin{equation}
    \hat{H}_{repulsion}=U \sum_{i} (1-n^{(\tau)}_{i})(1-n^{(\tau)}_{i+1})\,,
\end{equation} with $U\gg1$. This way the low-energy states are still given by states satisfying the local constraints, and the IR dynamics do not change.

\subsection{Emanant symmetry}
\label{ssec:sigmap}
The anyon chain Hamiltonian $\hat{H}_{\rm Fib}$ preserves the Fibonacci symmetry on the lattice level.
In addition to the Fibonacci symmetry, the field theory in the continuum limit has extra symmetries that emanate from the space group symmetry of the lattice \cite{Cheng:2022sgb,Seiberg:2023cdc}.

\begin{figure}[h]
\centering
\includegraphics[scale=0.4]{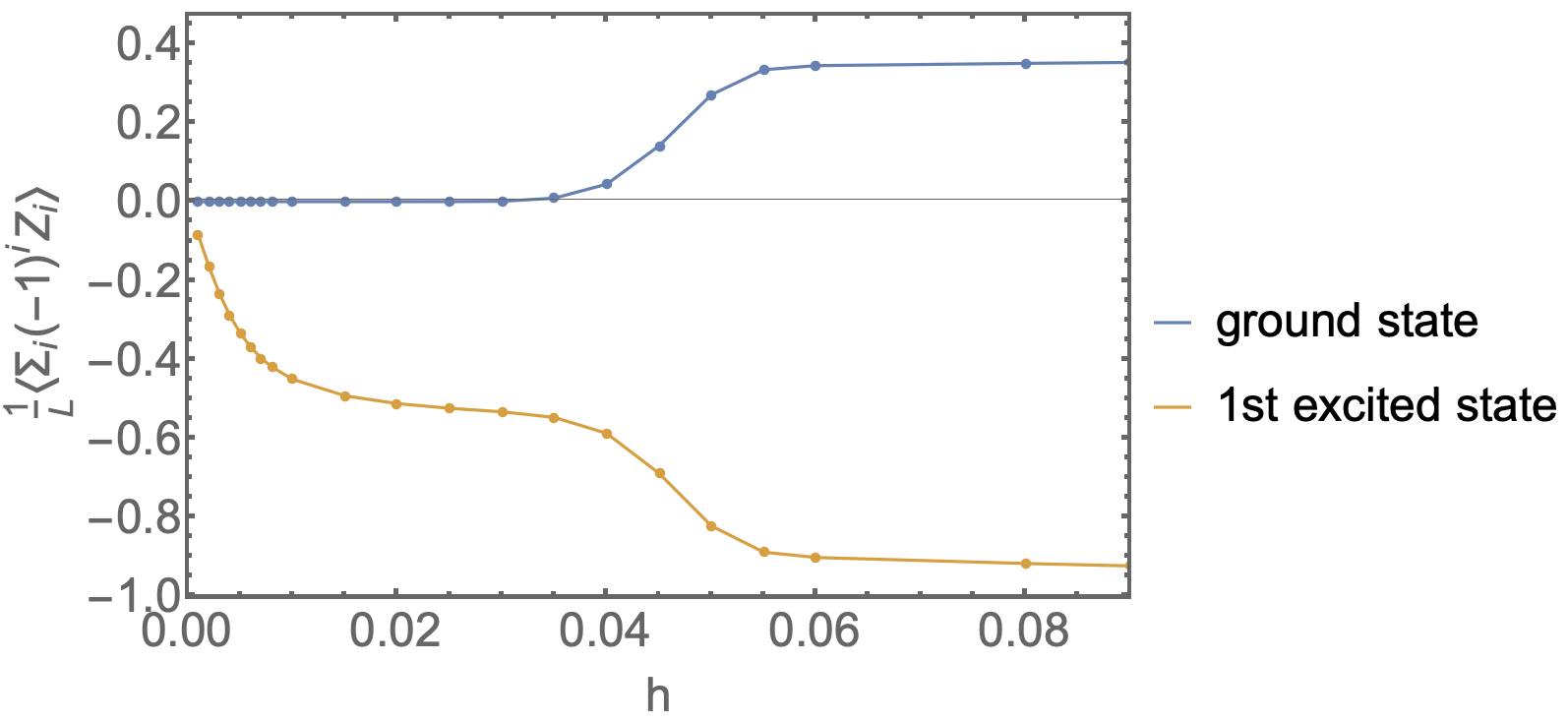}
\caption{The expectation value of $\langle \frac{1}{L}\sum_i(-1)^i Z_i\rangle$ as a function of $h$ for the two lowest lying states.}
\label{fig:sigmah2states}
\end{figure}

When $J>0$, the anyon chain model corresponds to the tri-critical Ising CFT.
The $\mathbb{Z}_2$ symmetry of the tricritical Ising model is not realized on-site, and instead maps to a translation by a single site in the lattice model. 
This can be seen explicitly by examining the exact diagonalization result of the anyon chain Hamiltonian reported in \cite{Feiguin:2006ydp}.
In particular, $\mathbb{Z}_2$ odd operators, like $\sigma'$, sit at the edge of the Brillouin zone, i.e. they have momentum $\pi$.
If we want to perturb our Hamiltonian by $\sigma'$ we therefore have
\begin{equation}
    \int dx \,\sigma'(x) \to \sum_i e^{i p x_i} O_i= \sum_i (-1)^i  O_i \,.
\end{equation}
Here, $O_i$ is a lattice operator to be determined later.
To understand the exact microscopic operator which gives rise to $ \sigma'$ we need to study the symmetries of the tri-critical Ising model in more detail. 
In particular, we recall that the action of the non-invertible Fibonacci line $W$ on local operators is given as in Table \ref{tab:Waction}. 
\begin{table}[h]
\centering
\begin{tabular}{|l|l|l|l|l|l|l|}
\hline
              & 1     & $\epsilon$                    & $\sigma$                      & $\sigma'$                     & $\epsilon'$                 & $\epsilon''$                \\ \hline
$(r, s)$  & (1,1) & (1,2)                         & (2,3)                         & (2,1)                         & (3,2)                       & (1,4)                       \\ \hline
$(h,\bar{h})$ & (0,0) & $(\frac{1}{10},\frac{1}{10})$ & $(\frac{3}{80},\frac{3}{80})$ & $(\frac{7}{16},\frac{7}{16})$ & $(\frac{3}{5},\frac{3}{5})$ & $(\frac{3}{2},\frac{3}{2})$ \\ \hline
$W$      & $\varphi$ & $-\varphi^{-1}$                   & $-\varphi^{-1}$                   & $\varphi$                         & $-\varphi^{-1}$                 & $\varphi$                       \\ \hline
\end{tabular}
\caption{The action of $W$ on conformal primaries of the tricritical Ising CFT. Here $\varphi=\frac{1}{2} \left(\sqrt{5}+1\right)$. }\label{tab:Waction}
\end{table}
The only two primaries invariant under $W$ (i.e. whose action is the same as on the identity), are the irrelevant $\mathbb{Z}_2$-even $\epsilon''$ and our target $ \sigma'$. In particular, since the Hamiltonian density is invariant under the Fibonacci symmetry it is natural to conjecture that
\begin{equation}
      \int dx \,\sigma'(x) \to \sum_i (-1)^i  \hat{H}_i \,.
\end{equation}
To verify this proposal, we study the deformed anyon chain
\begin{equation}
\hat{H}=\sum_i \hat{H}_i + h \sum_i (-1)^i \hat{H}_i\,,
\end{equation}
using DMRG.
First, to verify the explicit breaking of the $\mathbb{Z}_2$ symmetry, we compute the expectation value of the anti-ferromagnetic $\mathbb{Z}_2$ order parameter  $ \langle \frac{1}{L}\sum_i(-1)^i Z_i\rangle$ on the two lowest lying states ($Z$ is a shorthand notation for the Pauli matrix $\sigma_z$) as a function of the coupling constant $h$. The result at $L=80$ is summarized in Fig.~\ref{fig:sigmah2states}.
Notice $\langle \frac{1}{L}\sum_i(-1)^i Z_i\rangle$ corresponds the the expectation value of the leading operator odd under the $\mathbb{Z}_2$ symmetry, that is $\langle \sigma \rangle$.
At large enough $h$, both states have non-zero values of the order parameter which are \textbf{not} symmetric under a flip, signaling the explicit breaking of the $\mathbb{Z}_2$ symmetry. Furthermore these two states have almost the same energy, which follows from the non-trivial ground state degeneracy ensured by the non-invertible symmetry in the continuum limit \cite{Aasen:2016dop,Aasen:2020jwb,Copetti:2024rqj}. At small $h$, on the other hand, the true ground states have $\langle \frac{1}{L}\sum_i(-1)^i Z_i\rangle=0$, while the 1st excited state has non-zero $\langle \frac{1}{L}\sum_i(-1)^i Z_i\rangle$. Notice that at finite size, we can not have true spontaneous symmetry breaking; the true ground state is always the superposition of the two ground states.

\begin{figure}[h]
\centering
\includegraphics[scale=0.4]{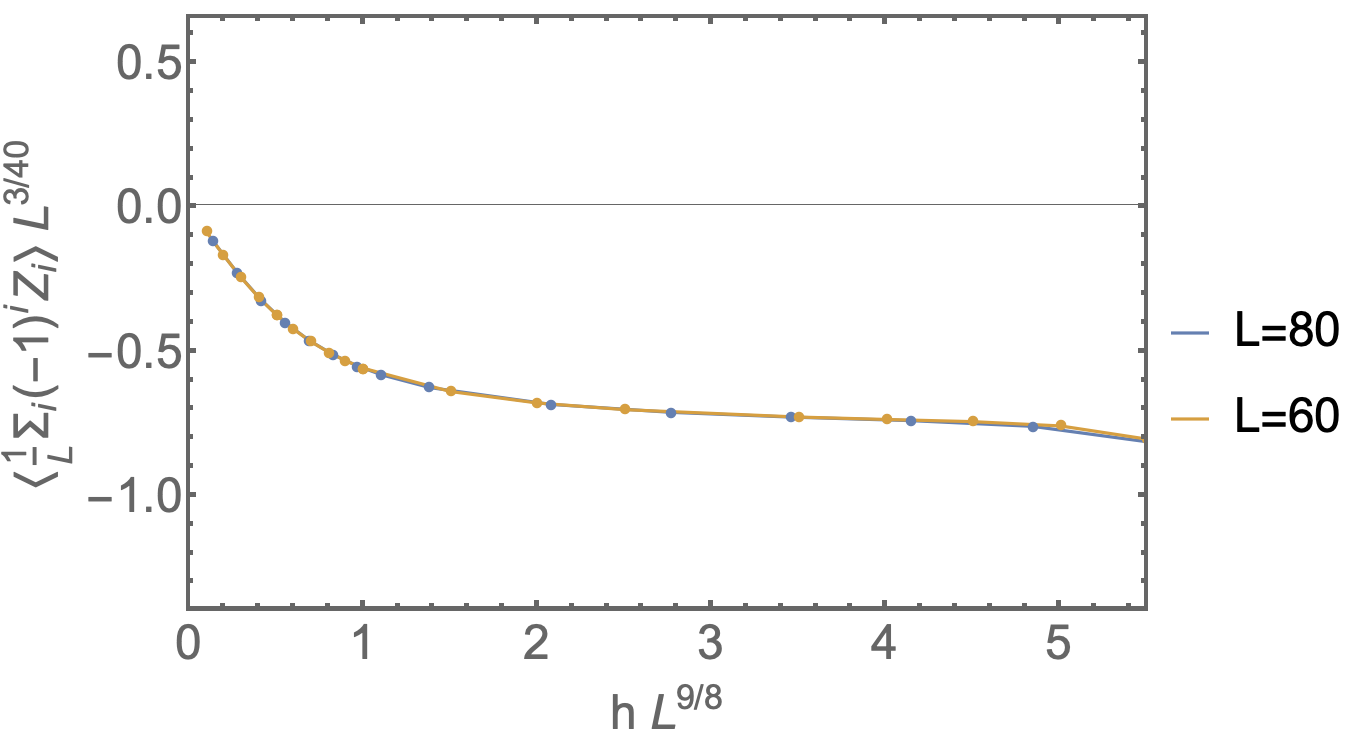}
\caption{Data collapsing of $\langle \frac{1}{L}\sum_i(-1)^i Z_i \rangle$ as a function of $h$ at two different lattice sizes. The expectation value is measured on the first excited state.}
\label{fig:datacollapsing1chain}
\end{figure}

We can also perform the finite-size scaling/data collapsing near the $h=0$ point, as shown in Fig.~\ref{fig:datacollapsing1chain}. We have used that $\Delta_{\sigma}=3/40$, and $\Delta_{\sigma'}=7/8$. The rescaled expectation values $\langle\sigma\rangle L^{\Delta_{\sigma}}$ as a function of $h L^{2-\Delta_{\sigma'}}$, is scale invariant and therefore independent of the lattice size $L$.

\begin{figure}[h]
\centering
\includegraphics[scale=0.4]{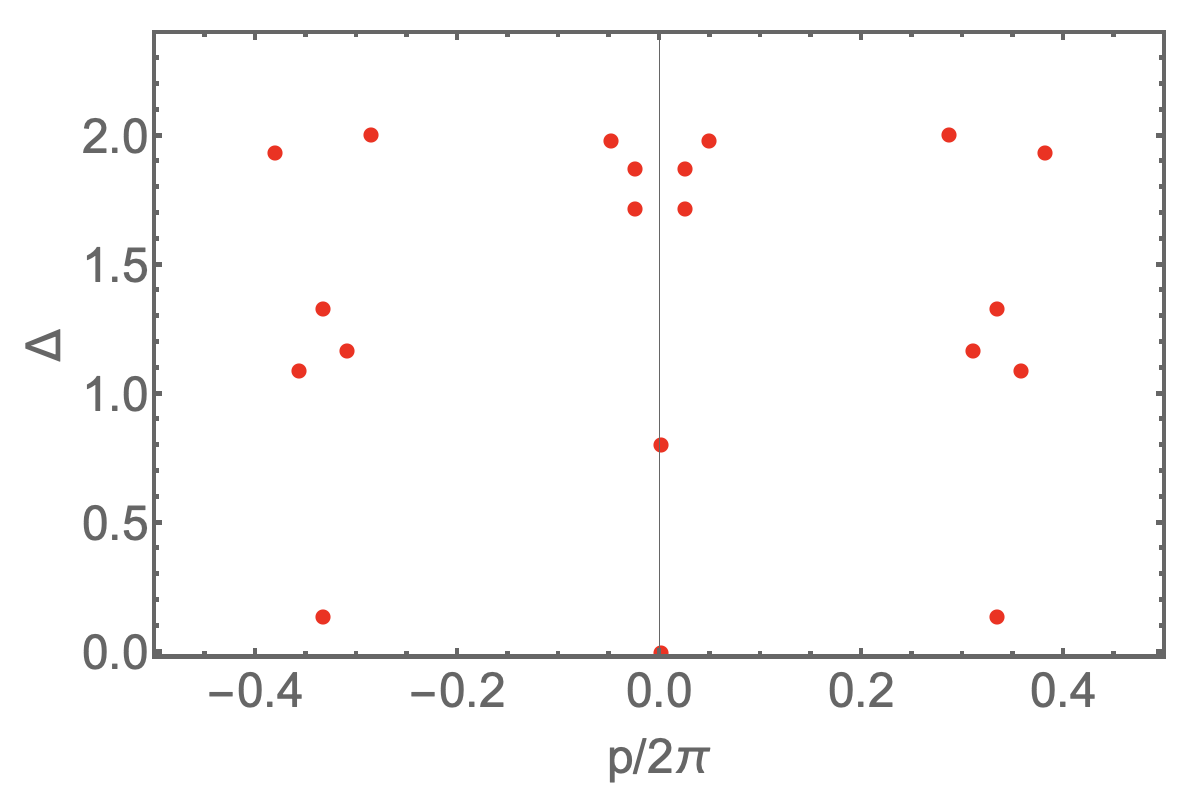}
\caption{The operator spectrum of the 3-state Potts model calculated using the anyon chain Hamiltonian. We have rescaled the leading excited state to $\Delta=2/15$.}
\label{fig:Pottsspectrum}
\end{figure}

When $J<0$, the anyon chain model corresponds to the 3-state Potts model CFT. The $\mathbb{Z}_3$ symmetry emanates from the lattice translation, while the charge conjugation symmetry comes from the reflection symmetry. Together, they form the non-abelian $S_3$ symmetry of the 3-state Potts model. We can also see this explicitly from the low-lying spectrum of the Hamiltonian, as reported in Fig.~\ref{fig:Pottsspectrum}. 
Up to finite-size corrections, the spectrum agrees with the CFT operator spectrum of the 3-state Potts model. One can also check that the $\mathbb{Z}_3$ charged operators carry momentum $\pm \frac{2\pi}{3}$.

\section{2-coupled anyon chains}
\label{sec:two}
\begin{figure}[h]
\centering
\includegraphics[scale=0.5]{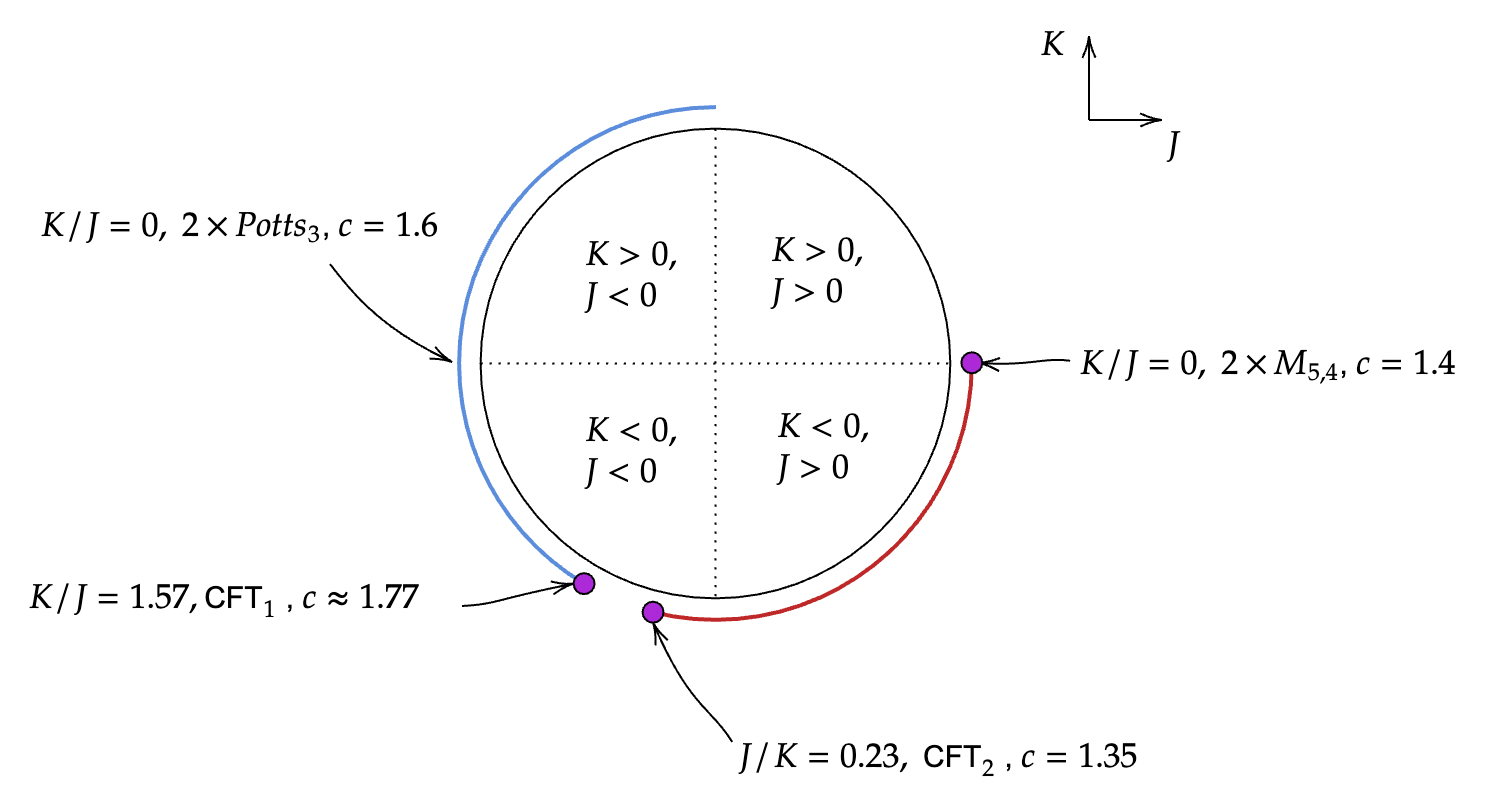}
\caption{The phase diagram of the 2-coupled anyon chain model. The red region is a pseudo-critical region, which is a weakly first-order phase transition. The blue region is a critical phase described by two decoupled Potts models. The unmarked regions are gapped. The two purple dots are CFTs. We do not know how to identify CFT$_1$ with any rational CFT. CFT$_2$ can be identified as the coset CFT $\frac{SU(2)_3 \times SU(2)_3}{SU(2)_6}$}
\label{fig:phasediagram}
\end{figure}

We now consider the following model of two coupled anyon chains
\begin{align}
\label{twochainHamil}
H= J \sum_i H^{(1)}_i + J \sum_i H^{(2)}_i +K \sum_{i} H^{(1)}_i H^{(2)}_i\,,
\end{align}
which manifestly preserves two copies of the Fibonacci symmetries. The $K$-term breaks independent one-site translations of the chains, but preserves the diagonal translational symmetry. For $J>0$ and $J<0$, this corresponds to breaking the $\mathbb{Z}_2^{(1)}\times \mathbb{Z}_2^{(2)}$  and the $S_3^{(1)}\times S_3^{(2)}$  symmetries to their diagonal subgroups respectively. The phase diagram is summarized in Fig.~\ref{fig:phasediagram}. For $J>0$ and $K=0$ we find that the two decoupled tricritical Ising CFTs ($c=1.4$) separate a weakly first order phase transition ($K<0$) in red from a gapped phase ($K>0$). Instead, for $J<0$ and $K=0$  we find a two decoupled 3-state Potts CFT ($c=1.6$), which in fact belongs to a conformal phase for a large range of values of $K$. Making $K$ sufficiently negative ($K/J\approx1.57$), we eventually find a non-trivial CFT with $c\approx1.77$ which we refer to as CFT$_1$ and which separates the $c=1.6$ conformal phase from a gapped regime. By making $K$ even more negative ($J/K=0.23$) we find a second non-trivial critical point, denoted CFT$_2$, which has $c=1.35$ and that separates the gapped phase from the weakly first order phase transition. In the remainder of this section we will discuss in detail each of these regions of the phase diagram.

\subsection{CFT$_1$}
It was noticed in \cite{Feiguin:2006ydp} that the single anyon chain model with negative coupling corresponds to the 3-state Potts model. Similarly to the tricritical Ising model, the $\mathbb{Z}_3$ symmetry of the 3-state Potts model is realized as a lattice translation by one or two sites. 
In this case, operators charged under the $\mathbb{Z}_3$ symmetry have momentum $\pm \frac{2\pi}{3}$.
Therefore, at the $J=-1$ and $K=0$ point, the 2-coupled anyon chain model~\eqref{twochainHamil} corresponds to two decoupled copies of the 3-state Potts CFTs.
Turning on the $K$ coupling does not lead to a gapped phase, and the model continues to be described by the decoupled Potts CFT with $c=1.6$.
This conformal phase is robust because the CFT that describes it has no relevant operators invariant under 
\begin{align}
    ( (\rm{Fibonacci})^2 \rtimes \mathbb{Z}_2)\times S_3\,,
\end{align}
which is the symmetry of the lattice model. The leading irrelevant corrections are described by the action \eqref{actionfield2}.

\begin{figure}[h]
\centering
\includegraphics[scale=0.4]{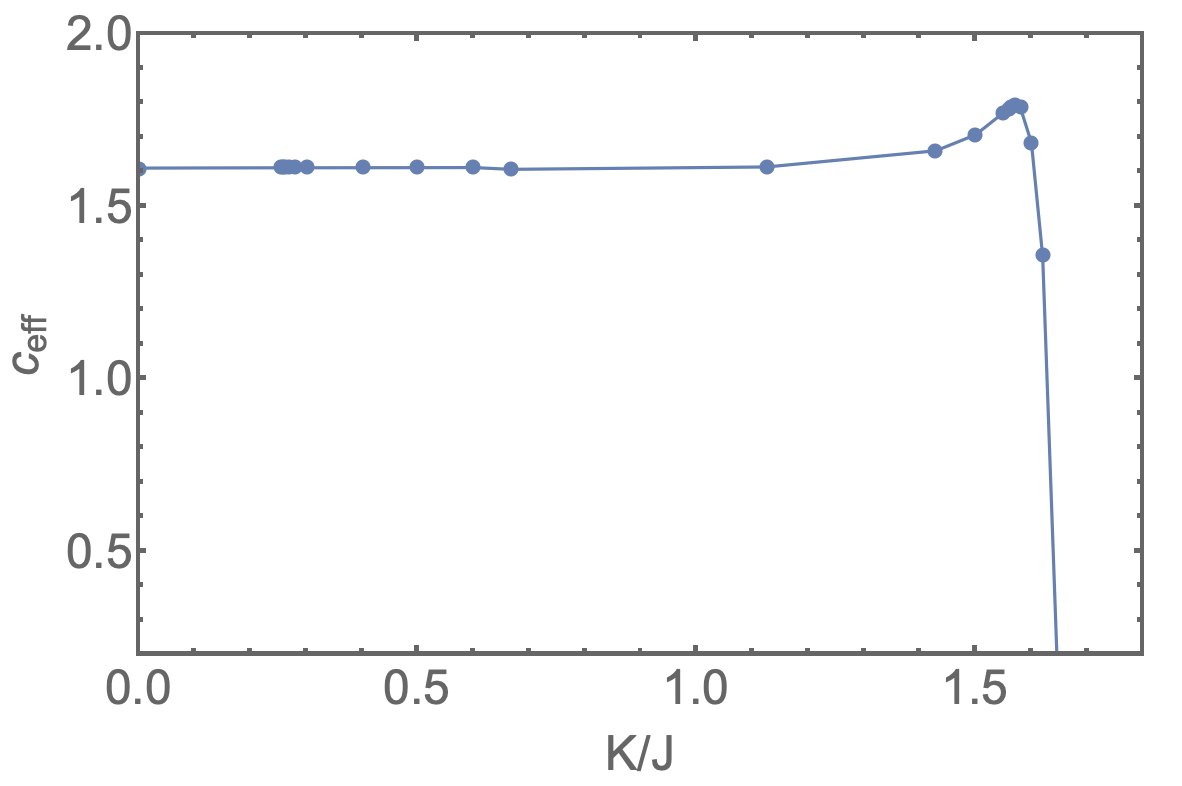}
\caption{The effective central charge as we vary the couplings near CFT$_1$.}
\label{fig:effectiveCCFT1}
\end{figure}

\begin{figure}[h]
\centering
\includegraphics[scale=0.32]{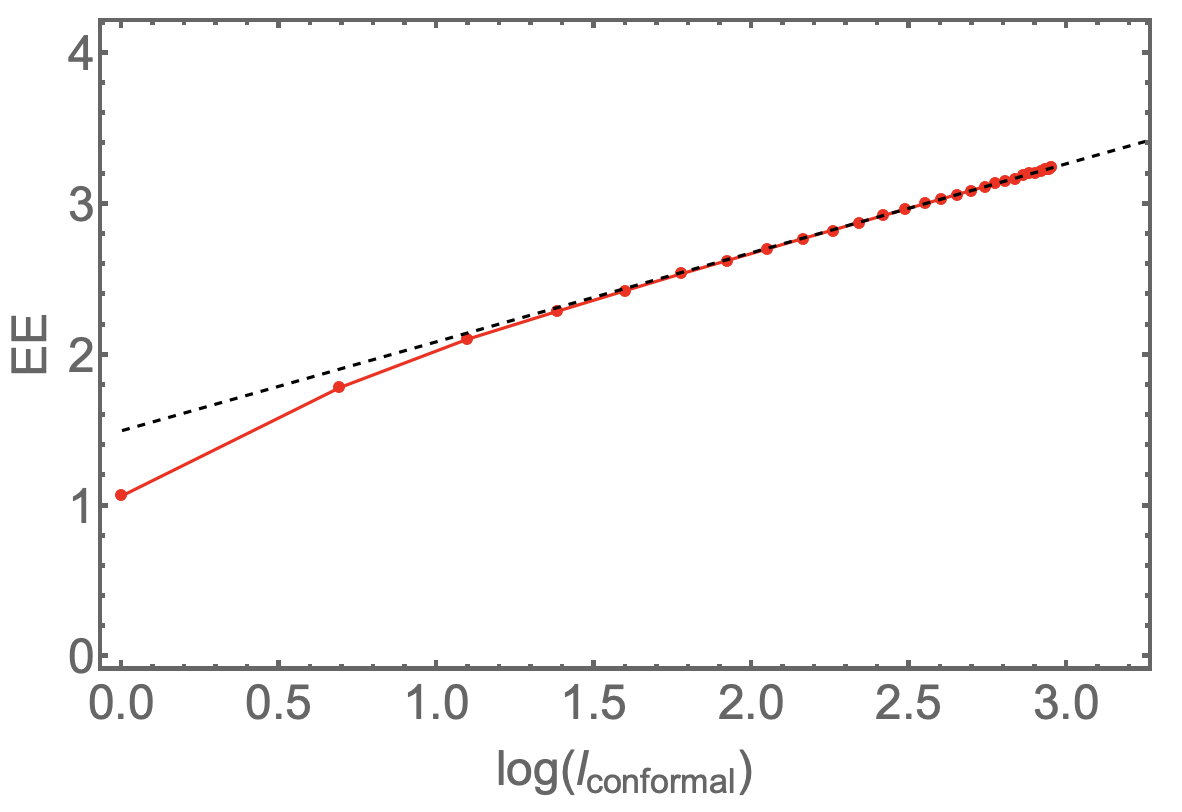}
\caption{The scaling of the entanglement entropy with the size of the cut interval for CFT$_1$. The data points are calculated using lattice size $L=60$ and maximum bond dimension $\chi=5000$.
}
\label{fig:EECFT1}
\end{figure}

\begin{figure}[h]
\centering
\includegraphics[scale=0.4]{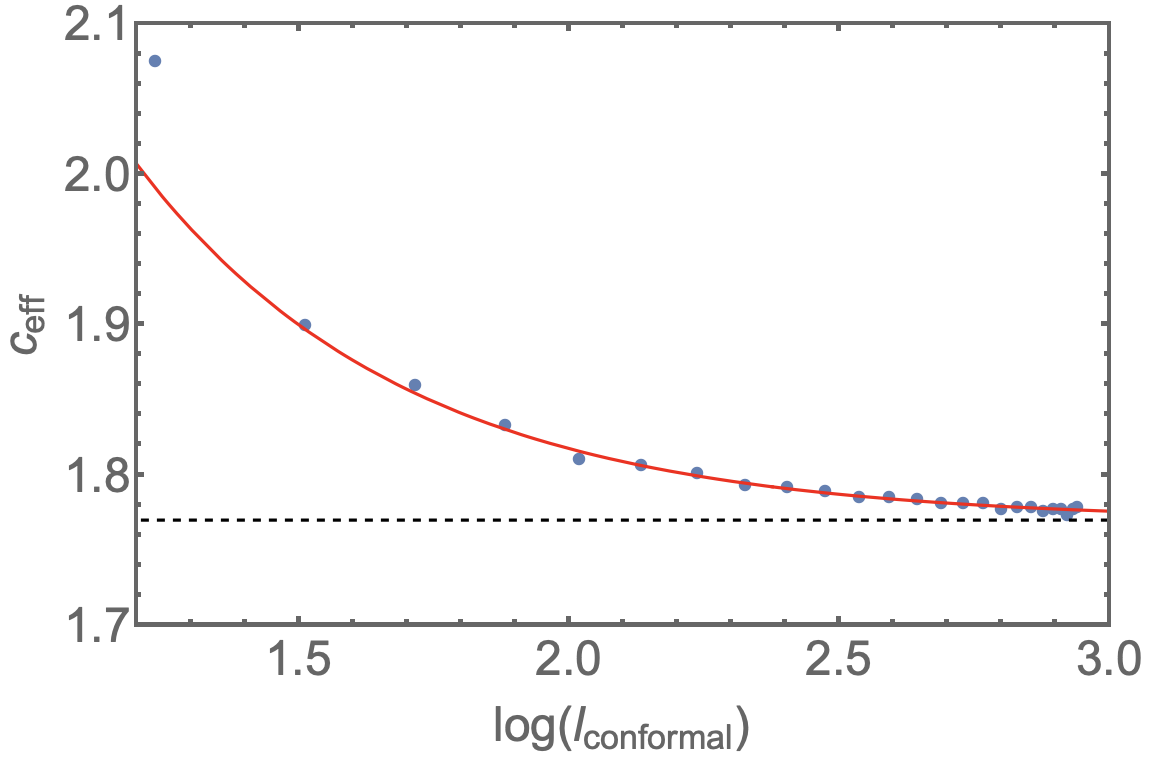}
\includegraphics[scale=0.42]{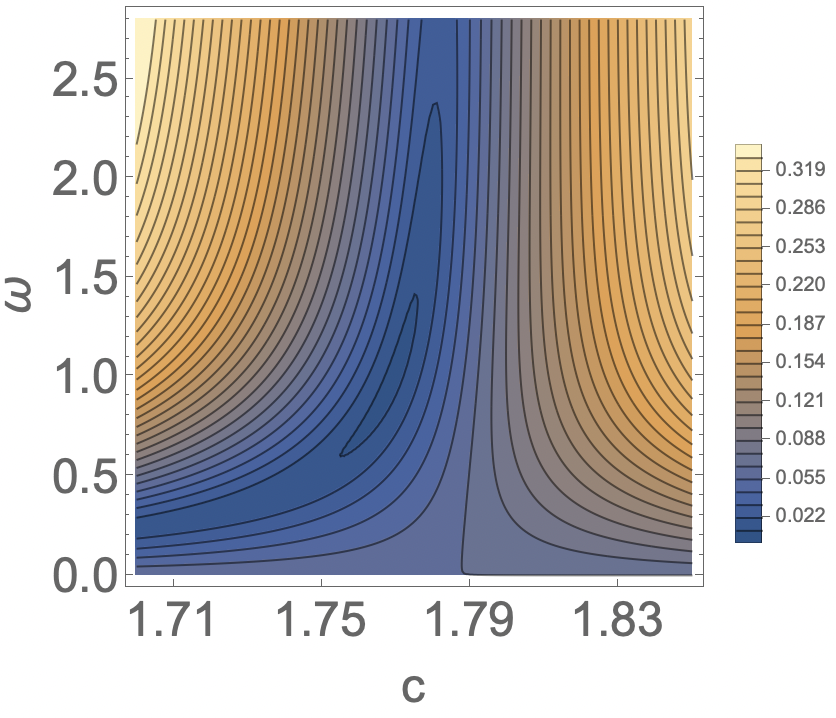}
\caption{Left: Fit of the effective central charge using \eqref{ceff}. 
The data points are obtained by doing a linear fit of four neighboring points in Fig.~\ref{fig:EECFT1} and extracting the slope. 
The five points near the boundary are discarded because they are likely to be affected by the UV details of the lattice.
Right: The contour plot of the fitting error.}
\label{fig:fittingerrorCFT1}
\end{figure}

\begin{figure}[h]
\centering
\includegraphics[scale=0.4]{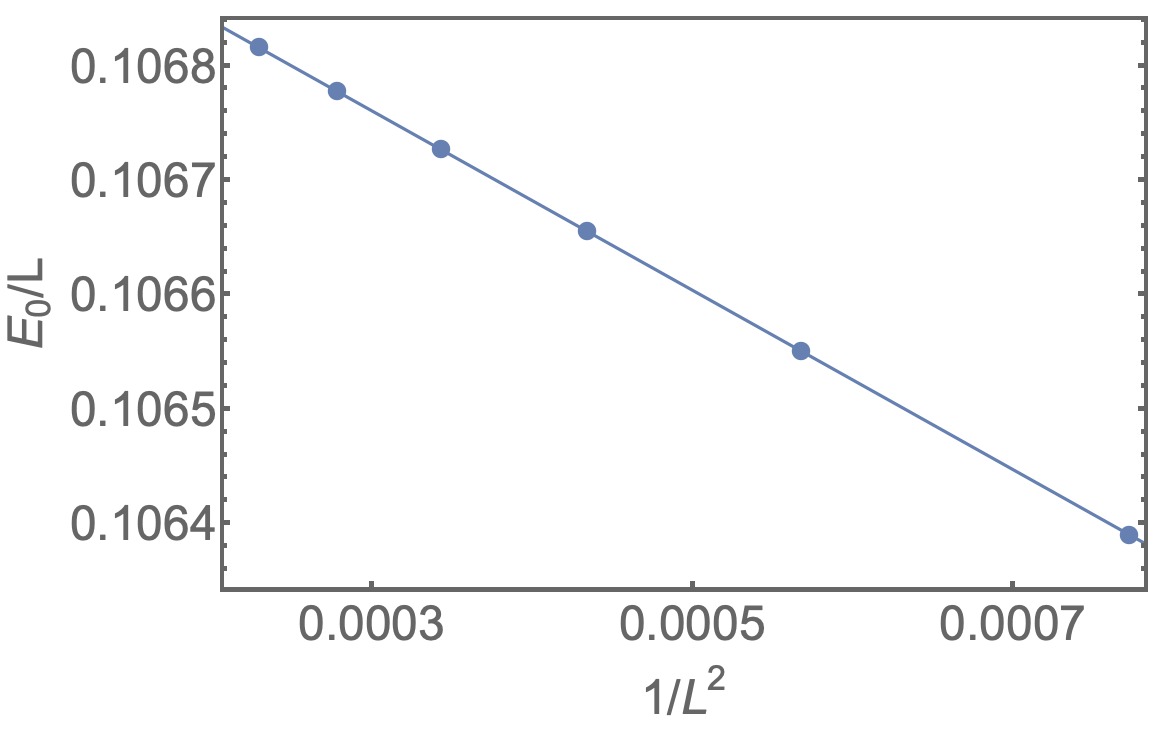}
\caption{The ground state energy as a function of the lattice size $L$. This is measured at the critical point CFT$_1$. The linear plot indicates that the system is a CFT with dynamical critical exponent $z=1$.}
\label{fig:dynamitcalexponentsCFT1}
\end{figure}

The critical point denoted as CFT$_1$ in Fig.~\ref{fig:phasediagram} corresponds to a second order phase transition point separating the critical phase (two decoupled Potts CFTs) and a gapped phase.
We can study the effective central charge around this critical point. This is summarized in Fig.~\ref{fig:effectiveCCFT1}. 
To get the central charge of CFT$_1$, we study the entanglement entropy between a line segment with length $l$ and the result of the spin chain with length $L-l$ at the critical point $K/J \approx1.57$. This is reported in Fig.~\ref{fig:EECFT1}.
We fit the effective central charge using the expression 
\begin{align}
\label{ceff}
    c_{\rm eff}(l)=c+e^{-2\omega t}, \quad {\rm with}\quad t= \log(l_{\rm conformal}),
\end{align}
and find
\begin{equation}
    c_1\approx1.77\,.
\end{equation}
We justify and discuss the formula \eqref{ceff}  in Appendix~\ref{app:finitesize}.
The quality of the fitting can be analyzed using the contour plot of the fitting residues, as shown in Fig.~\ref{fig:fittingerrorCFT1}.
To show that this critical point is an actual CFT, we need to check the dynamical critical exponent, which can be calculated by studying how the ground state energy scales with the lattice size. This is summarized in Fig.~\ref{fig:dynamitcalexponentsCFT1}. One can see that the ground state energy follows \eqref{energyscaling}, as expected for a Lorentzian symmetric ($z$=1) conformal field theory~\footnote{In a general critical theory with $z\neq1$ one instead expects to observe a scaling $L^{-z}$ for the leading non-extensive part of the ground state energy, see for example~\cite{Wang_2022}.}. 

\begin{figure}[h]
\centering
\includegraphics[scale=0.4]{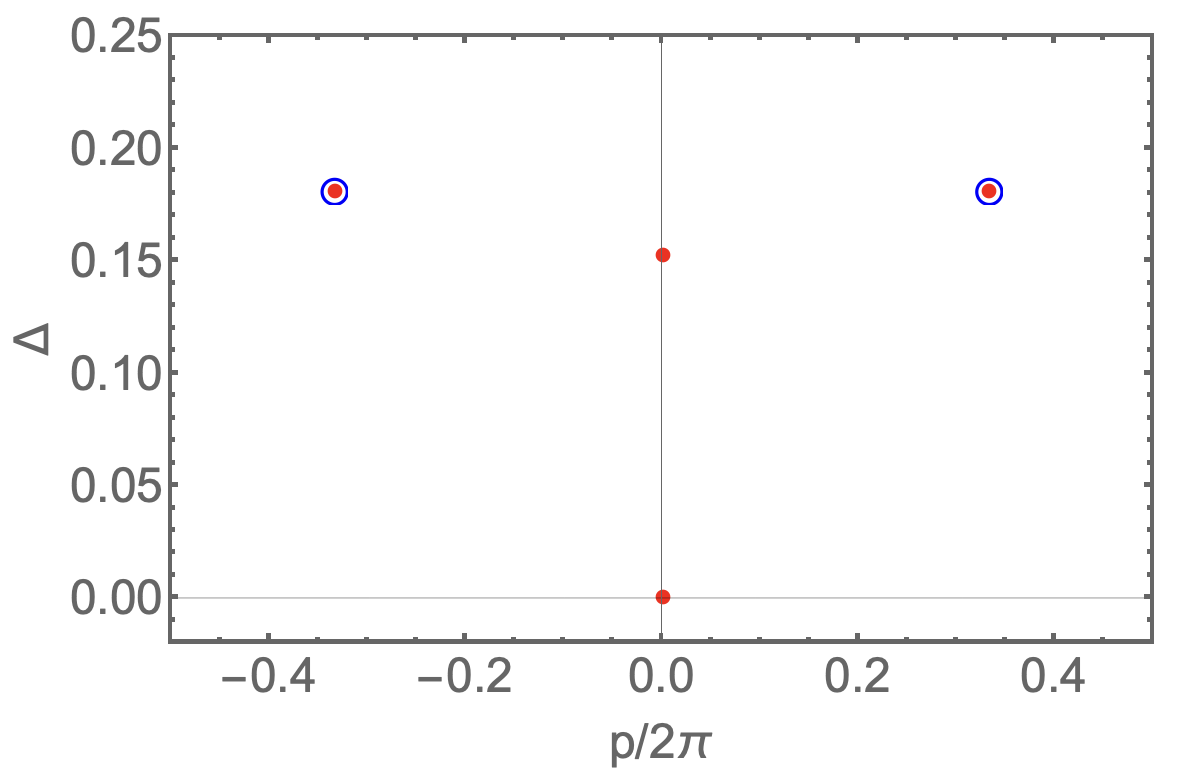}
\caption{The operator spectrum of CFT$_1$. The red dots/blue circles indicate operators that are even/odd under the permutation symmetry $\mathbb{Z}_2^{(p)}$.}
\label{fig:spectrumCFT1}
\end{figure}

We can also study the spectrum of this CFT. 
We have already measured the central charge $c$ in Fig.~\ref{fig:EECFT1}. 
The result in Fig.~\ref{fig:dynamitcalexponentsCFT1} then allows us to measure $E_{\infty}$ and the effective ``speed of light'' $v$ that appears in equation \eqref{energyscaling}. 
Together they give us the relation between the energy spectrum and the scaling dimension of operators.
They are summarized in Fig.~\ref{fig:spectrumCFT1}.
Remarkably, we sometimes find degeneracy between states which are even and odd under the $\mathbb{Z}_2^{(p)}$ symmetry.
The degeneracy is obvious in the decoupled IR theory.
Given any two operators $\mathcal{O}$ and $\mathcal{O}'$ in a given copy of the Potts model, we can always use them to build two degenerate states in the tensor product theory:
\begin{equation}
    |\mathcal{O}_1,\mathcal{O}'_2\rangle\,, \qquad |\mathcal{O}'_1,\mathcal{O}_2\rangle\,,
\end{equation}
which can further be combined into eigenstates of the permutation $\mathbb{Z}_2^{(p)}$ symmetry by taking the sum and difference. 
This also makes it clear that some states will not be degenerate, which is the case when we take $\mathcal{O}'=\mathcal{O}$. 
In fact, we claim that the twofold degeneracies are enforced by the non-abelian nature of the non-invertible symmetry
\begin{equation}
    (\textrm{Fibonacci} \times \textrm{Fibonacci}) \rtimes \mathbb{Z}_2^{(p)}\,,
\end{equation}
in full analogy with the existence of two dimensional representations for the invertible non-abelian symmetry $D_4=(\mathbb{Z}_2\times\mathbb{Z}_2)\rtimes \mathbb{Z}_2$, where the rightmost $\mathbb{Z}_2$ acts by permuting the actions of the first two.\footnote{Consider the two-dimensional Hilbert space spanned $|\sigma \rangle\otimes |1\rangle$ and $|1 \rangle\otimes |\sigma\rangle$, the symmetry acts on it as 
\begin{align}
    W^{(1)}= \left(
\begin{array}{cc}
 -\varphi^{-1} & 0 \\
 0 & \varphi  \\
\end{array}
\right),
\quad  W^{(2)}= \left(
\begin{array}{cc}
 \varphi  & 0 \\
 0 & -\varphi^{-1} \\
\end{array}
\right), 
\quad \mathbb{Z}_2^{(p)}=\left(
\begin{array}{cc}
 0 & 1 \\
 1 & 0 \\
\end{array}
\right).
\end{align}
One can easily see that the two-dimensional Hilbert space does not have an invariant subspace under the action of the above matrices.
The two states, therefore, should have the same energy.
The precise mathematical statement is related to the G-extension of modular tensor categories, see for example~\cite{delaney2019fusionrulespermutationextensions,Benjamin:2025knd}. 
We will leave a rigorous study of this symmetry to future work.} The upshot of this analysis is that CFT$_1$ is a candidate for a UV completion of two decoupled 3-state Potts models whose leading irrelevant interactions are described by \eqref{actionfield2}.

\subsection{The pseudo critical phase}
\begin{figure}[h]
\centering
\includegraphics[scale=0.4]{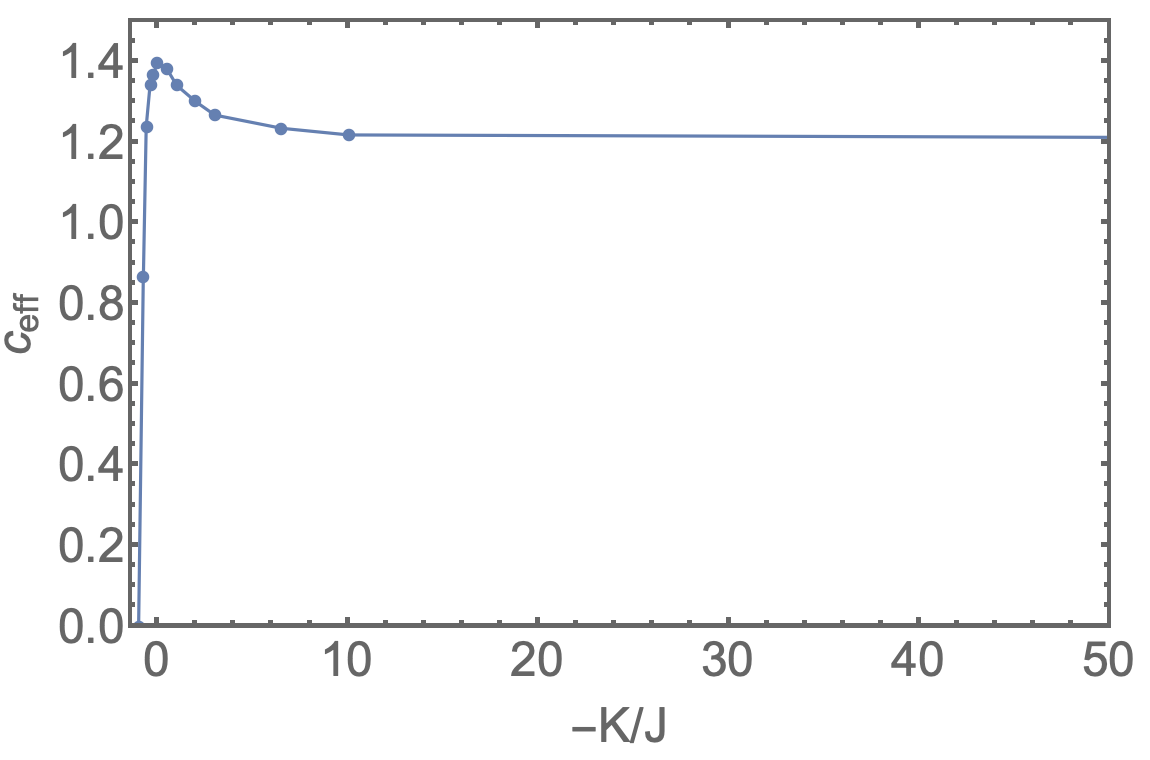}
\caption{The effective central charge as a function of $-K/J$. This is a measurement at $L=80$.}
\label{fig:effectiveC2chain}
\end{figure}

\begin{figure}[h]
\centering
\includegraphics[scale=0.3]{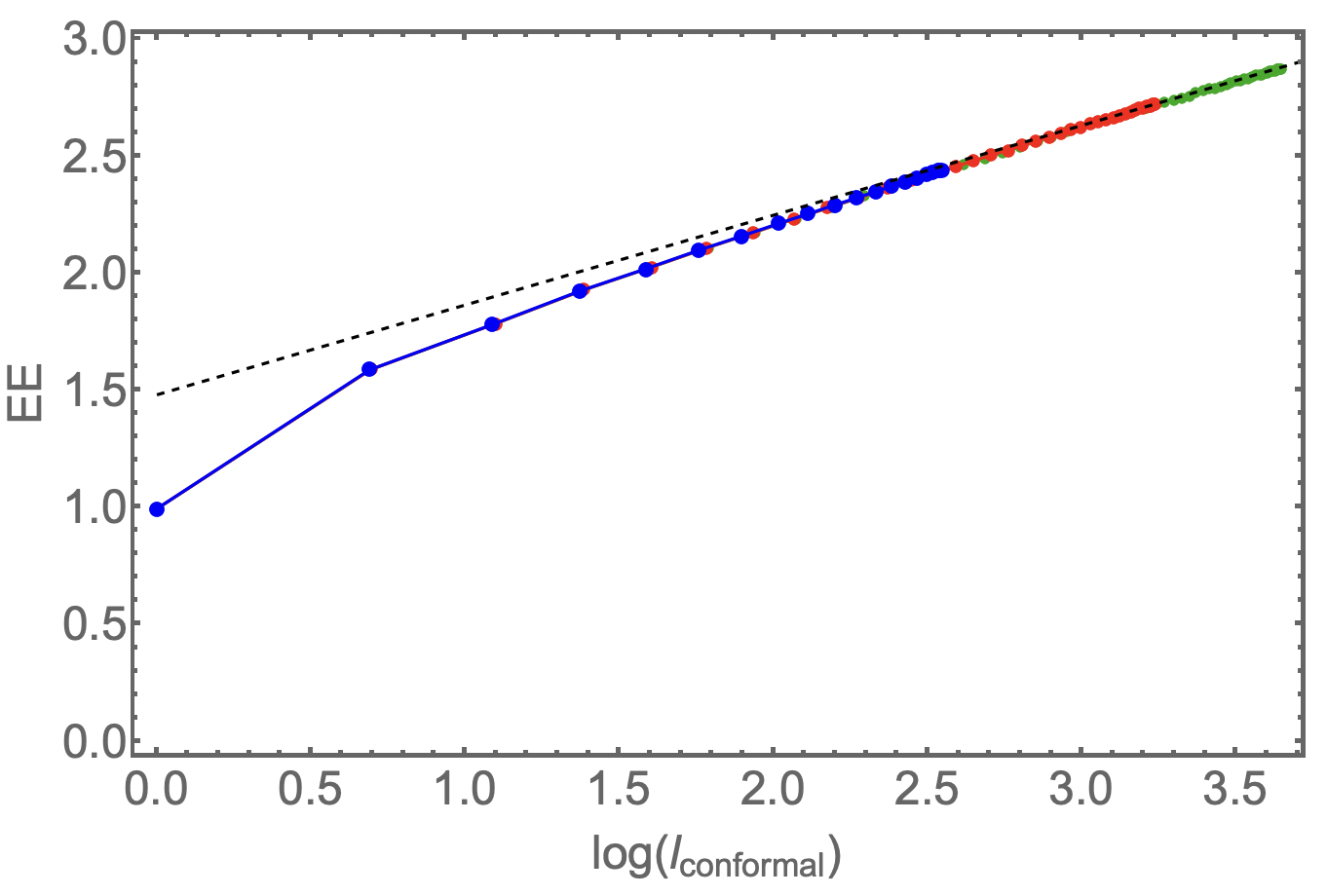}
\includegraphics[scale=0.35]{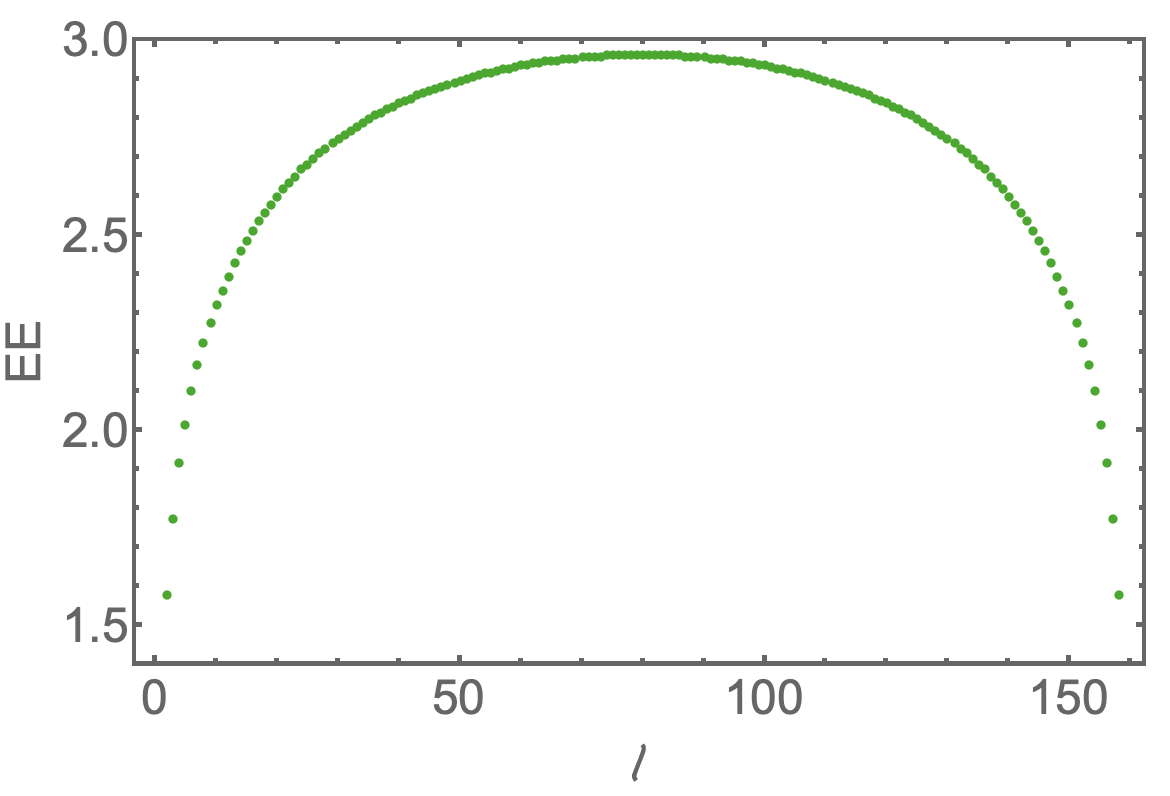}
\caption{Entanglement entropy at $K/J=\infty$. The blue, red, and green points in the left figure correspond to $L=40,80$ and $120$, respectively. The overlapping of the data points is due to the conformal symmetry.
The dashed line corresponds to $EE=1.151\times\frac{\log{(l_{\rm conformal})}}{3}  + 1.48$. The right figure is calculated at $L=120$. The maximum bond dimension we used is $\chi=3200.$}
\label{fig:EElc3l}
\end{figure}

We now move our attention to the region near $J=1,K=0$. 
At the $J=1$, $K=0$ point, the theory becomes two-decoupled anyon chains; the continuum limit can therefore be described as two decoupled tri-critical Ising models. The interaction 
\begin{align}
\sum_{i} H^{(1)}_i H^{(2)}_i \sim \int dx ~h^{(1)}(x)h^{(2)}(x) \sim \int dp~ h^{(1)}(-p)  h^{(2)}(p)\,,
\end{align}
which couples to the $K$ term in the Hamiltonian \eqref{twochainHamil} corresponds to the deformation by the operator  
\begin{align}\label{operatormixing}
   \mathcal{O} =\sigma'^{(1)}\sigma'^{(2)}+ \big( \epsilon''^{(1)}+\epsilon''^{(2)} \big)+\epsilon''^{(1)} \epsilon''^{(2)} +\ldots.
\end{align}
The coefficients in front of the continuum operators depend on the details of the lattice model and are unfixed. The $\sigma'^{(1)}\sigma'^{(2)}$ operator is a relevant operator under the lattice symmetry, which in this case becomes 
\begin{align}
    ( (\rm{Fibonacci})^2 \rtimes \mathbb{Z}_2)\times \mathbb{Z}_2\,.
\end{align}
We can study the effective central charge $c_{\textrm{eff}}$ as we vary $K$: the results are summarized in Fig.~\ref{fig:effectiveC2chain} and suggest that at large negative values of $K/J$, $c_{\textrm{eff}}$ remains constant.
We can therefore focus on the $K/J=-\infty$ point to discuss the pseudo-critical behavior. The entanglement entropy is reported in Fig.~\ref{fig:EElc3l}.
Picking the data points with $\log(l_{\rm conformal})>3.25$ and fitting against equation~\eqref{centralchargeL}, we get the central charge   
\begin{align}
    c_{\rm eff} =1.151\pm 0.001.
\end{align}
The error reported here is the standard error of the linear-fit parameter. 
We can also study the dynamical critical exponent.
\begin{figure}[h]
\centering
\includegraphics[scale=0.45]{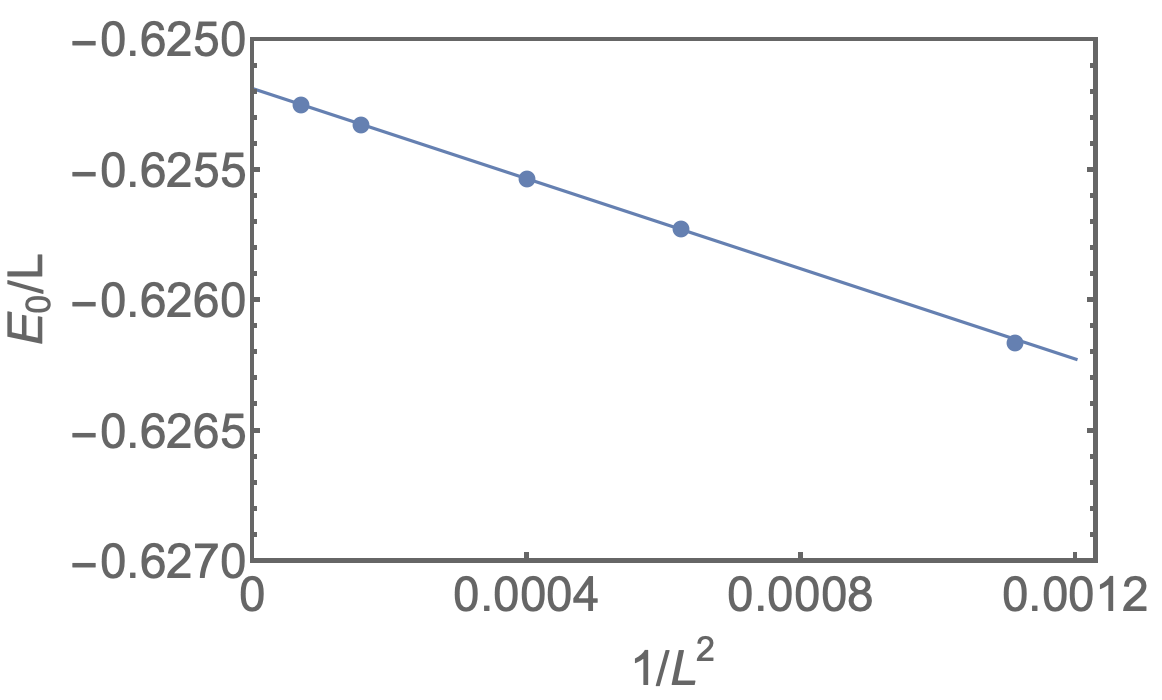}
\caption{The ground state energy density $E_0/L$ versus the (squared) lattice size $L^2$. 
The data is calculated at $K/J=\infty$.}
\label{fig:Eglc3l}
\end{figure}
As one can see in Fig.~\ref{fig:Eglc3l}, the ground state energies follow the scaling relation~\eqref{energyscaling}. We can even study the spectrum of this ``CFT'' like  we did for CFT$_1$. The energy eigenvalues are summarized in Fig.~\ref{2chainSPEC}.
\begin{figure}[htbp]
\centering
\includegraphics[scale=0.45]{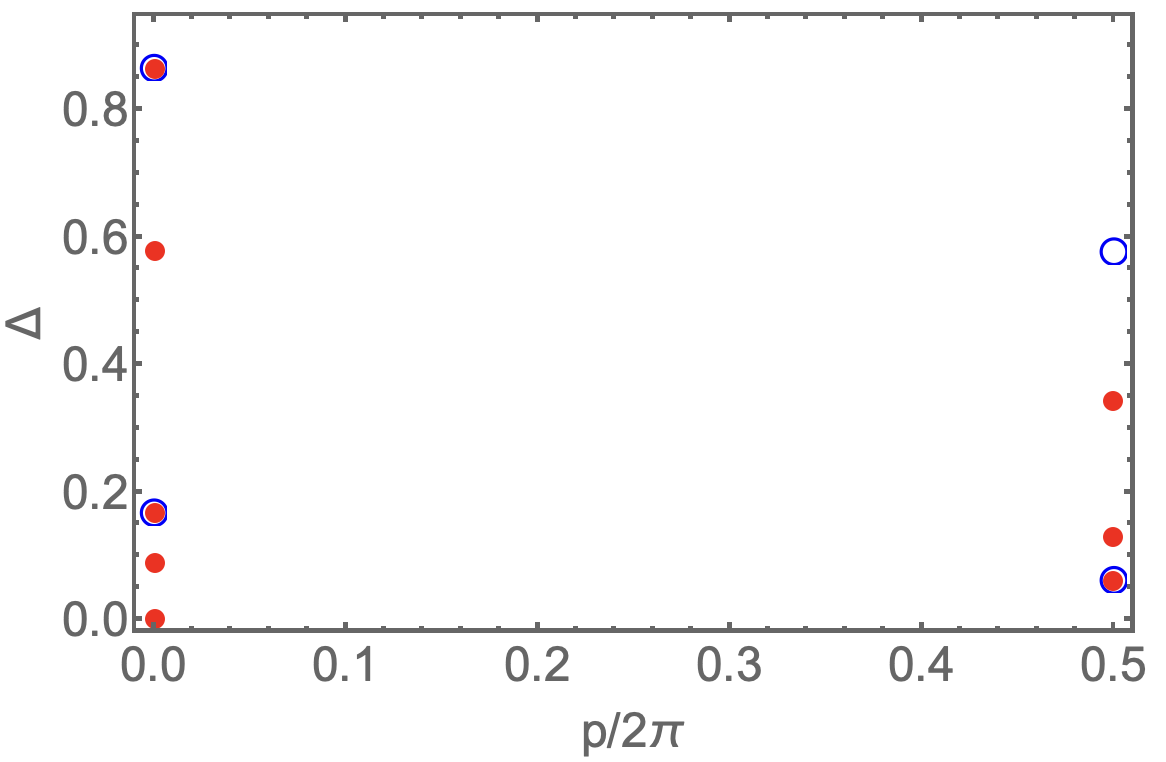}
\caption{The spectrum of the two coupled chains measured at $K/J=\infty$. The size of the system is $L=40$.
The system has a $\mathbb{Z}_2^{(p)}$ permutation symmetry which exchanges the states on the two chains. The red dots/blue circles indicate states that are even/odd under the $\mathbb{Z}_2^{(p)}$ symmetry.
In addition to $\mathbb{Z}_2^{(p)}$, the system has an emanant $\mathbb{Z}_2$ symmetry, which corresponds to translating the lattice by one site. 
This can be seen from the fact that the low-energy states cluster around $p=0$ and $p=\pi$. }
\label{2chainSPEC}
\end{figure}

\begin{figure}[h]
\centering
\includegraphics[scale=0.38]{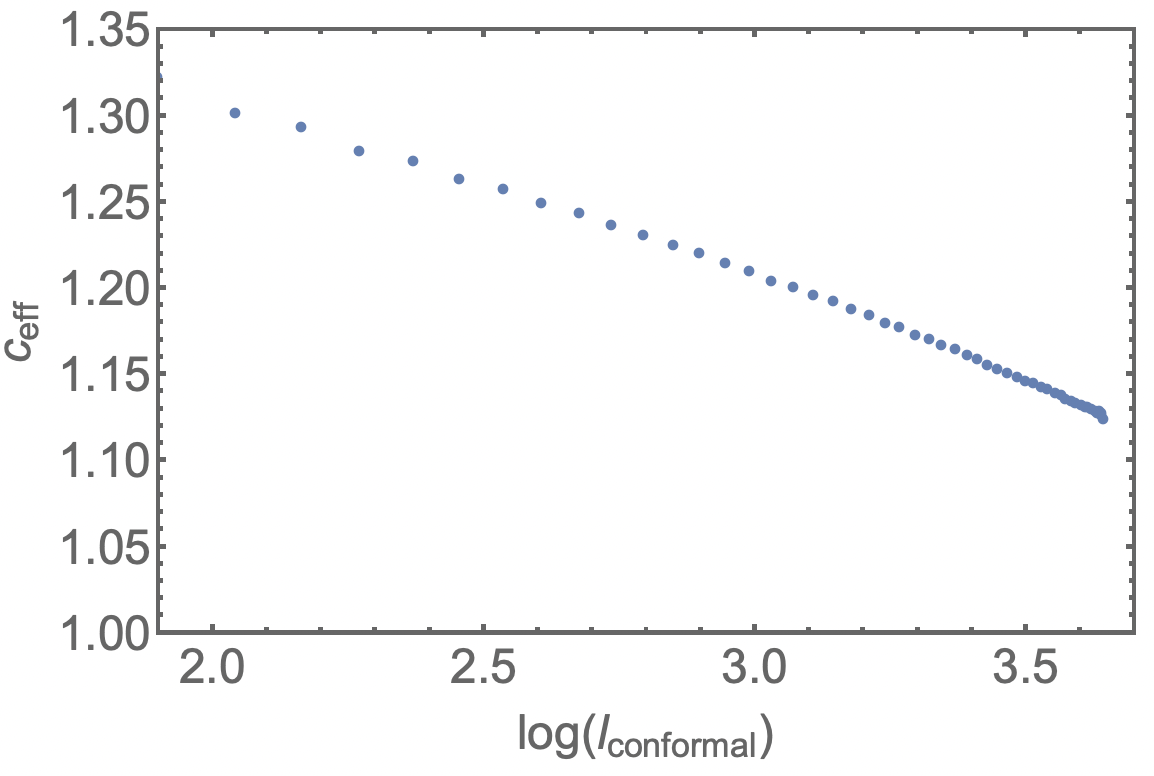}
\includegraphics[scale=0.42]{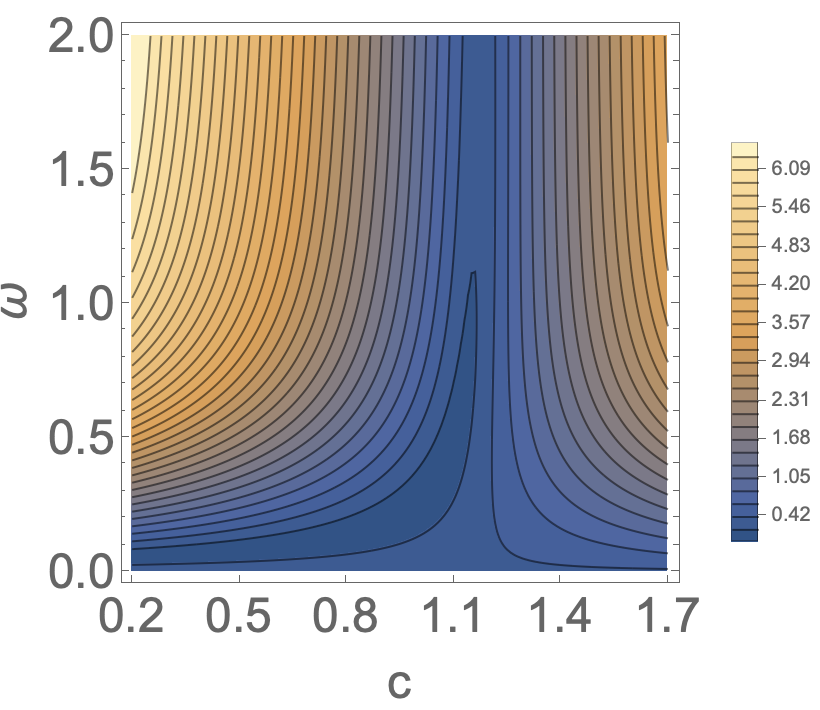}
\caption{Left: The effective central charge as a function of $l_{\rm conformal}$ for the pseudo critical phase. 
Right: The contour plot of the residue of the fittings using \eqref{ceff}.}
\label{fig:ceff_lc_PC}
\end{figure}

All these results seem to suggest that we have a critical phase and furthermore that it is a relativistic CFT with dynamical critical exponent $z=1$.
What gives away that we are actually in the presence of pseudo-critical behavior is a detailed study of how the effective central charge changes as we measure it for different system sizes, as shown in Fig.~\ref{fig:ceff_lc_PC}.
Notice that the effective central charge never reaches a constant value, which means we cannot be really be at a critical point. In fact, it is possible to show that the $K/J\to \infty$ limit of our Hamiltonian \eqref{twochainHamil} is integrable and gapped, but with a large correlation length. The authors of \cite{Blakeney:2025ext} were able to map our Hamiltonian to a single integrable anyon chain for the fusion category $\rm{Fib}^2$ which exhibits a large correlation length $\xi \approx155$. The integrability of the Hamiltonian is proven via its connection to the Temperley-Lieb algebra. The Temperley-Lieb algebra also played a key role in previous work in the Euclidean transfer-matrix formalism \cite{Fendley:2008fp,Vernier_2014} which had also observed this phenomenon: the infinite coupling limit of two $Q-$state Potts models in the loop formulation is described by a single $Q^2$-state Potts model, which is gapped for $Q>2$, as is the case for the tricritical Ising model, which has $Q=\varphi\approx2.62$.\footnote{We thank Hubert Saleur for first pointing this out and Jesper Jacobsen for illuminating discussions on this point.} In fact, this means that our weakly first order phase transition corresponds to the walking RG behavior in the vicinity of a complex CFT \cite{Gorbenko:2018dtm,Gorbenko:2018ncu}.

Regardless of the weakly first order nature of the phase transition, there is a puzzle in interpreting our numerical result. 
If we deform the decoupled tricritical Ising model by the term 
\begin{align}
    g \int d^2x~\sigma'^{(1)}\sigma'^{(2)},
\end{align}
the physics should not depend on the sign of the coupling, since we can always shift the sign of the coupling by the field redefinition
\begin{align}
    \sigma'^{(1)}\rightarrow -\sigma'^{(1)},
\end{align}
which is a symmetry of the undeformed theory.
Our lattice model clearly depends on the sign of the coupling constant $K$. 
This indicates that the irrelevant operators in~\eqref{operatormixing} are dangerously irrelevant operators.
\begin{figure}[htbp]
\centering
\includegraphics[scale=0.32]{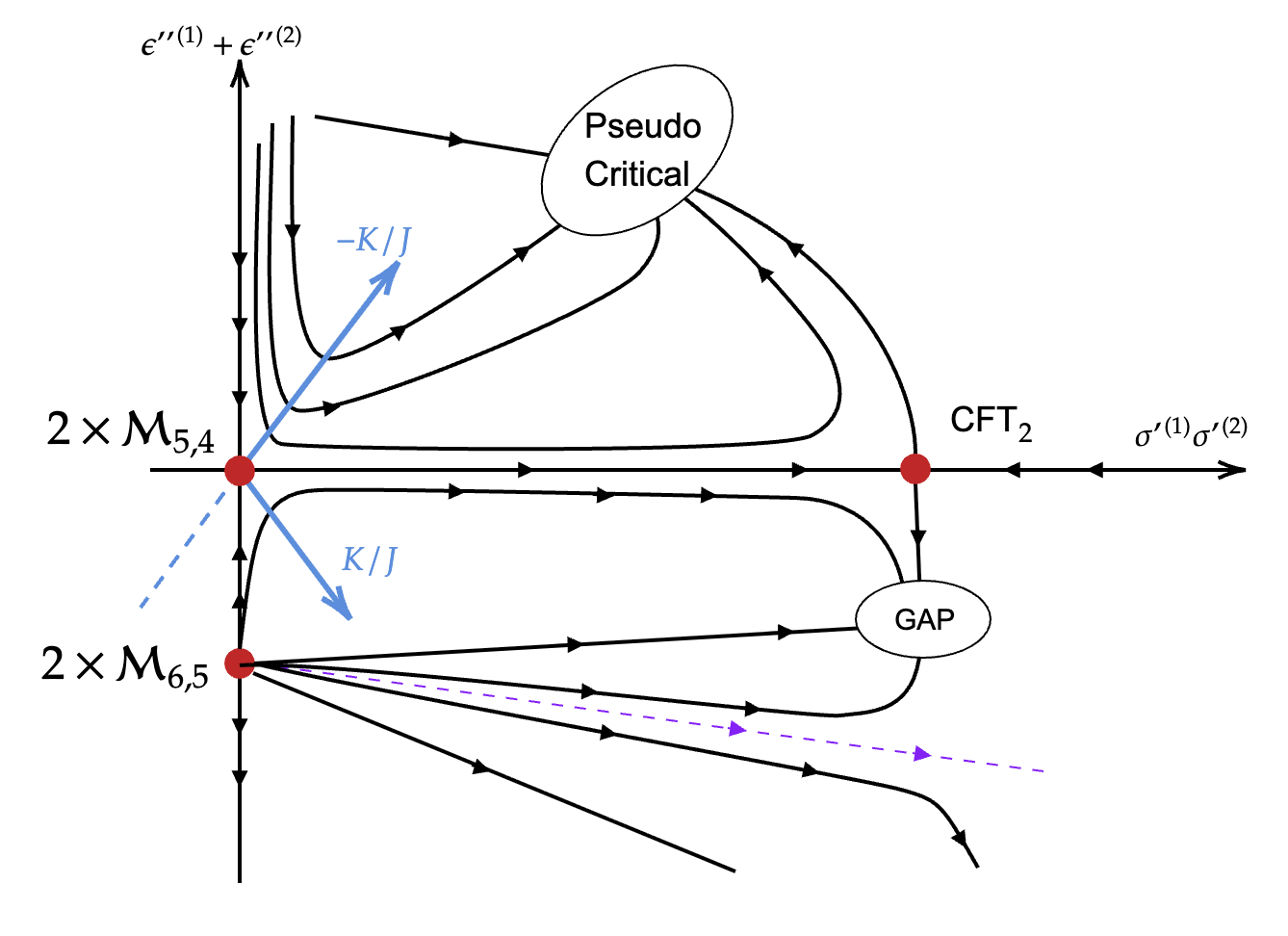}
\includegraphics[scale=0.3]{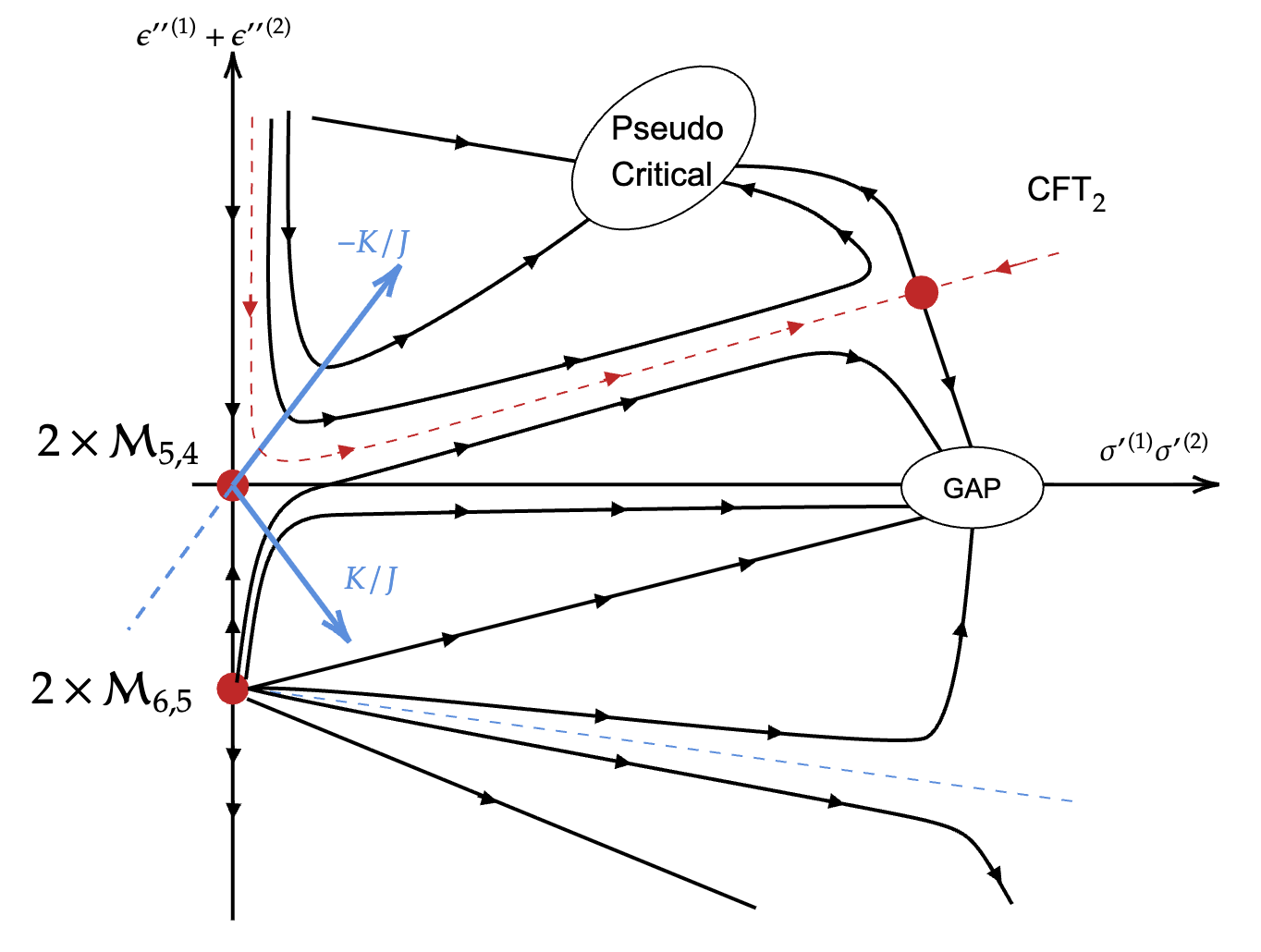}
\caption{Two possible scenarios of the RG flows of the 2-chain model near $J=1$ and $K=0$. The blue arrow corresponds to the $K/J$. Notice the plots are symmetrical along the vertical axis. }
\label{RG2chain}
\end{figure}
Combining our numerical result with various earlier works, we now have a fairly complete understanding of $\mathcal{C}$-preserving RG flows from the decoupled tricritical Ising fixed point ($2\times \mathcal{M}_{5,4}$).
The flows are summarized in Fig.~\ref{RG2chain}. 
The ($2\times \mathcal{M}_{5,4}$) fixed point has one relevant direction.
Depending on the sign of $K/J$, the RG flow can reach either the pseudo-critical phase or the gapped phase.
This is conveniently explained by the existence of another CFT, which we now denote as CFT$_2$. In the next section, we will show that  CFT$_2$ can also be found in our phase diagram and confirm that there are indeed RG flows from CFT$_2$ to the gapped phase and pseudo-critical phase. 
However, the precise location of CFT$_2$ in the RG diagram remains obscure.
In fact, the pure $\sigma'^{(1)}\sigma'^{(2)}$ deformation was claimed to lead to a gapped phase in \cite{LECLAIR1998523}, even though an explicit construction of the S-matrix was skipped in their work.
This supports the second scenario in Fig.~\ref{RG2chain}. 
This scenario, however, indicates that the lattice model with small $-K/J$ should be in the gapped phase, instead of the pseudo-critical phase. 
We do not see such a behavior in our DMRG simulation, see Fig.~\ref{fig:effectiveC2chain}. 
This leaves the possibility that the actual RG flows follow the first scenario in Fig.~\ref{RG2chain}.
Notice also that the $\epsilon''\equiv\phi_{(3,1)}$ deformation is the leading irrelevant operator that controls the (inverted) Zamolodchikov flows, i.e. we expect to be able to reach a fully unstable decoupled tetracritical Ising fixed point ($2\times \mathcal{M}_{6,5}$) by tuning an additional parameter in the lattice model.\footnote{The $\sigma'^{(1)}\sigma'^{(2)}$ deformation of ($2\times \mathcal{M}_{6,5}$) also leads to a massive phase, this is again discussed in~\cite{LECLAIR1998523}. }

\subsection{CFT$_2$: the coset CFT $(SU(2)_3 \times SU(2)_3)/SU(2)_6$}

As we mentioned previously, we need an extra CFT with the symmetry
\begin{align}
    ( (\rm{Fibonacci})^2 \rtimes \mathbb{Z}_2)\times \mathbb{Z}_2\,,
\end{align}
to explain the RG flow.
A natural candidate is the coset \cite{Goddard:1984vk,Goddard:1986ee,DiFrancesco:1997nk}\footnote{We thank Yifan Wang for first suggesting to us this CFT as a candidate.}
\begin{equation}
\label{su23coset}
    \frac{SU(2)_3 \times SU(2)_3}{SU(2)_6}\,.
\end{equation}
We provide a brief review of $SU(2)$ coset CFTs in Appendix \ref{app:coset}. For our analysis, we simply need to note that
the coset \eqref{su23coset} has an invertible center $\mathbb{Z}_2$ symmetry as well as a complex conjugation $\mathbb{Z}_2$ which swaps the two factors in the numerator, along with two Fibonacci lines $1_{L,R}$ that satisfy the truncated $SU(2)_3$ tensor product $1_{L,R}\times1_{L,R}=0_{L,R}+1_{L,R}$, ensuring the existence of some twofold degeneracies as discussed above. 
This theory has $c=1.35$ and furthermore contains a relevant operator with scaling dimension $\Delta=1.5$ which is invariant under the full non-invertible symmetry, as we show in Appendix \ref{app:RG}.

\begin{figure}[ht]
\centering
\includegraphics[scale=0.4]{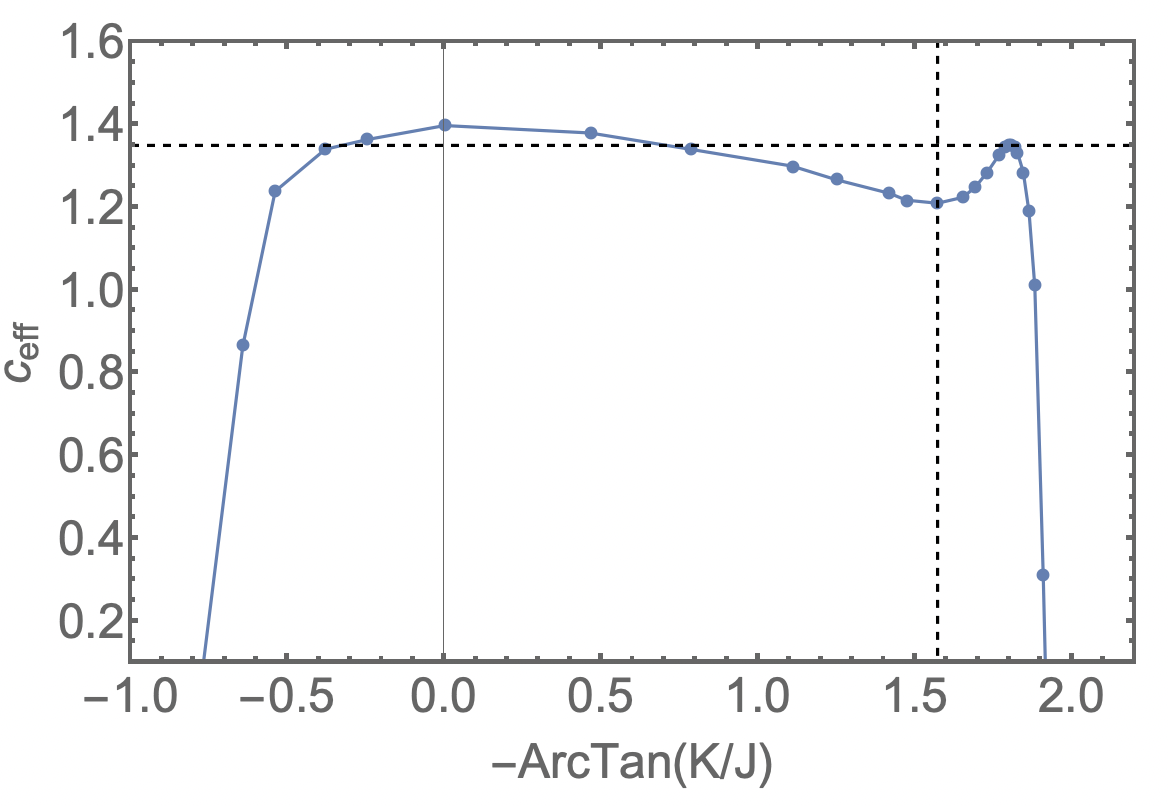}
\caption{The effective central charge as a function of $-{\rm Arctan}(K/J)$. 
This is a measurement at $L=80$, with maximum bond dimension $\chi=2000$.
The region with $-{\rm Arctan}(K/J)<\pi/2$ is the same as what was plotted in Fig.~\ref{fig:effectiveC2chain}. The full vertical line corresponds to $K=0$ while the dashed vertical line corresponds to $K=-\infty$. Finally the dashed horizontal line sits at $c_{\textrm{eff}}=1.35$ and is neatly saturated by the point we identify as CFT$_2$. }
\label{fig:effectiveC2chaincftp}
\end{figure}

\begin{figure}[ht]
\centering
\includegraphics[scale=0.4]{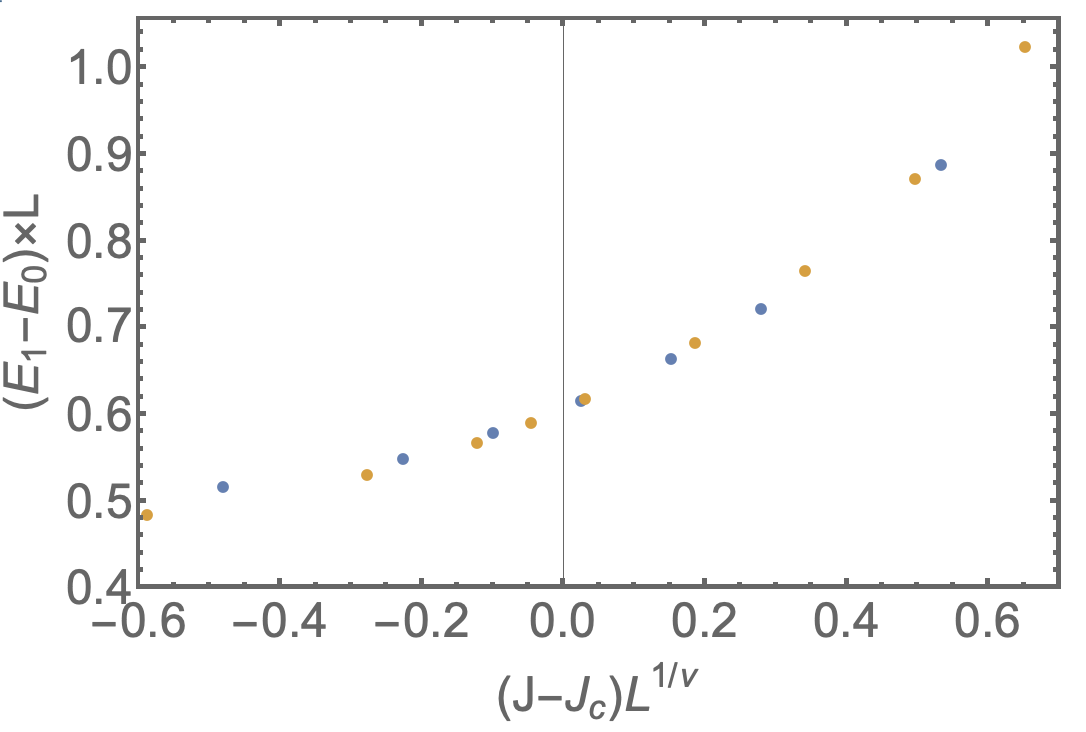}
\caption{The ``data collapsing'' analysis of the critical point corresponding to CFT$'$. Here $E_0$ is the ground state energy and $E_1$ is the energy of the leading state in the $Z_2^{(p)}$ odd sector. We take $1/\nu=0.5$ and $J_c/K=0.236$. The blue/orange dots correspond to data measured at $L=40$ and $L=60$, respectively.}
\label{fig:dataclpsCFTp}
\end{figure}

As anticipated above, this CFT also appears in our phase diagram.
In Fig.~\ref{fig:effectiveC2chain}, we studied our lattice model \eqref{twochainHamil} up to $K/J=-\infty$. 
This corresponds to fixing $J=0$ and $K=-1$.
We can continue to study the region with $J/K>0$, as shown in Fig.~\ref{fig:effectiveC2chaincftp}. 
There exists a local maximum of the effective central charge $c\approx 1.35$ at $J/K\approx 0.23$ which we identify as CFT$_2$. Notice that CFT$_2$ is a critical point separating a pseudo-critical phase and a gapped phase, as predicted in Fig.~\ref{RG2chain}.\footnote{This effective central charge plot closely resembles Figure 5 of \cite{Fendley:2008fp}, which studies a similar model in the euclidean formalism.}
We can use the standard ``data collapsing'' method to study the critical exponents of this critical point, as shown in Fig.~\ref{fig:dataclpsCFTp}. These turn out to be consistent with the existence of the relevant singlet of dimension $\Delta=1.5$ identified in Appendix \ref{app:RG}.

\section{3-coupled anyon chains}
\label{sec:three}
In this section, we will identify an IR fixed point of the three coupled tricritical Ising models using conformal perturbation theory at small but finite coupling. This fixed point was conjectured to be irrational in \cite{Chang:2018iay}. Then, using DMRG we identify a stable conformal phase that we conjecture to be described precisely by this putative irrational CFT.
\subsection{Naive conformal perturbation theory}
We begin by arguing for the existence of a reliable fixed point with $N\geq3$ coupled tricritical Ising models. We recall the formal action
\begin{equation}
    S=\sum_{a=1}^{N} S_{\rm tri-Ising}^{(a)} + K \int d^2x \sum_{a\neq b}\sigma'^{(a)}\sigma'^{(b)}\,.
\end{equation}
To study this model in conformal perturbation theory (CPT), we need to understand the structure of the self-OPE of the perturbation. For the $N=2$ case we have
\begin{equation}
    \sigma'^{(1)}\sigma'^{(2)} \times \sigma'^{(1)}\sigma'^{(2)} \sim 1 + (\epsilon''^{\,(1)}+ \epsilon''^{\,(2)})\,,
\end{equation}
which does not contain any relevant operators (and in particular does not contain $\sigma'^{(1)}\sigma'^{(2)}$ itself). Using the standard result for the beta function in conformal perturbation theory \cite{Zamolodchikov:1987ti}
\begin{equation}
    \beta_i= (2-\Delta_i)g^i -\pi \sum_{j,k}C_{jk}^i g^j g^k+O(g^3)\,,
\end{equation}
it becomes clear that there is no one-loop fixed point. Note that for a systematic CPT analysis, we need a small parameter $\epsilon$ such that $\Delta=2-\epsilon$ and that UV correlation functions can be computed in an expansion in $\epsilon$. For a small but finite value of epsilon (such as $1/4$ in our case), we can only use the leading order formulas (see for example \cite{Komargodski:2016auf,Rong:2023owx} for other examples of such "uncontrolled" conformal perturbation theory). Fortunately, for $N\geq3$ a one-loop fixed point is actually available. Indeed, noting that
\begin{equation}
    \sigma'^{(i)} \sigma'^{(j)} \times \sigma'^{(j)} \sigma'^{(k)} \sim \sigma'^{(i)} \sigma'^{(k)} + \textrm{irrelevant}\,,
\end{equation}
allows us to obtain
\begin{equation}
    \beta_K= \frac{K}{4}-\pi \frac{6}{\sqrt{3}^3}K^2=0\,,
\end{equation}
where 6 is a combinatorial factor for the OPE coefficient and the factors of $\sqrt{3}$ ensure unit normalization of the perturbing operator. This beta function has a UV fixed point at $K=0$ and an IR fixed point at $K^*=\sqrt{3}/(8\pi) \approx 0.07$, which is numerically small, suggesting the fixed point might be trustworthy. Once the one-loop fixed point is established, it is straightforward to use Cardy's c-theorem sum-rule \cite{Zamolodchikov:1986gt,Cardy:1988tj}
\begin{equation}
    \Delta c= \frac{\epsilon^3}{(C_{KK}^K)^2}\,
\end{equation}
to obtain an estimate for the IR central charge 
\begin{equation}\label{perturbativecentral}
    c_{IR}\approx2.088\,,
\end{equation}
which is remarkably close to the UV value, as is expected from the cubic nature of the central charge correction. Below, we will find a stable conformal phase with a central charge compatible with this value.  
\subsection{Numerical analysis}
\begin{figure}[ht]
\centering
\includegraphics[scale=0.4]{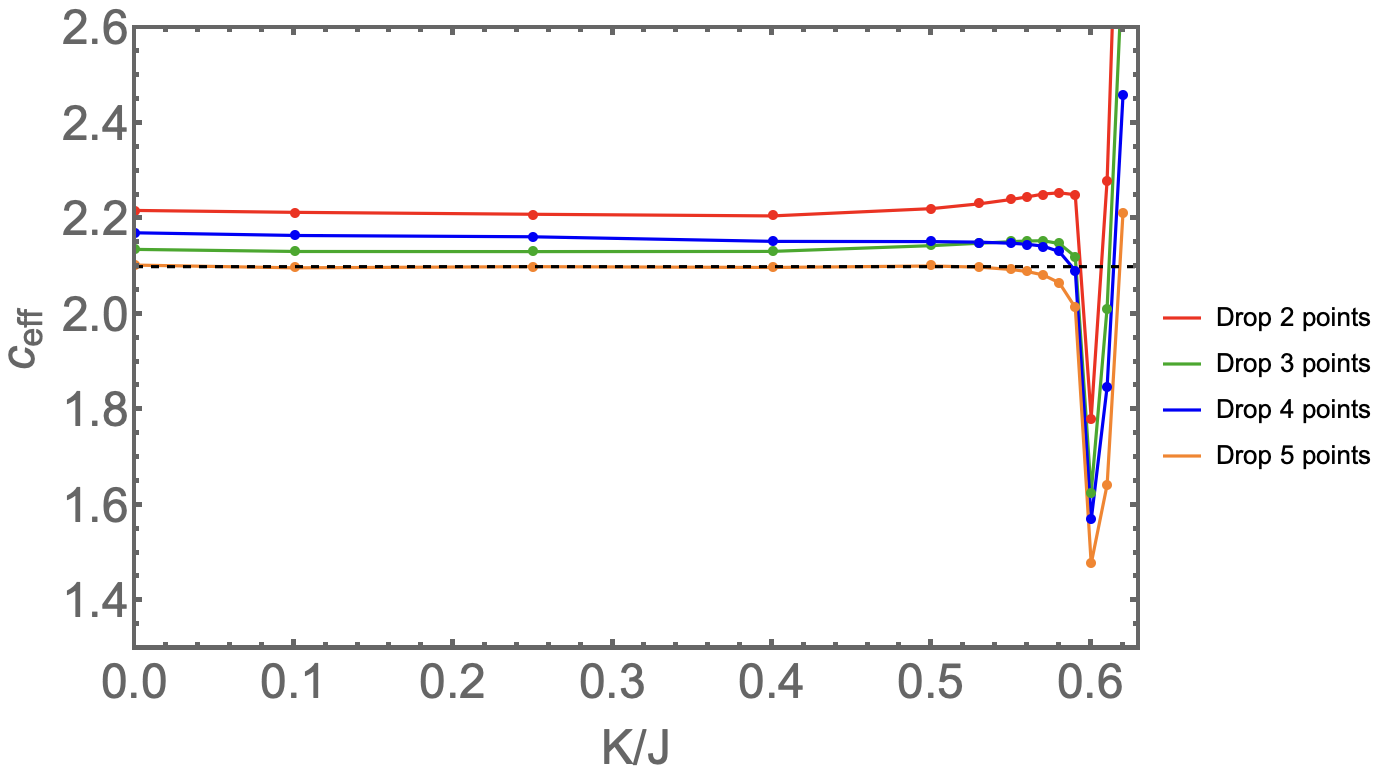}
\caption{The effective central charge of the three-chain model in the $K/J>0$ region. We observe a conformal phase, which corresponds to the proposed irrational conformal field theory.
The results are obtained using $L=20$ and maximum bond dimension $\chi=1500$. }
\label{fig:3chaingapless}
\end{figure}

\begin{figure}[ht]
\centering
\includegraphics[scale=0.4]{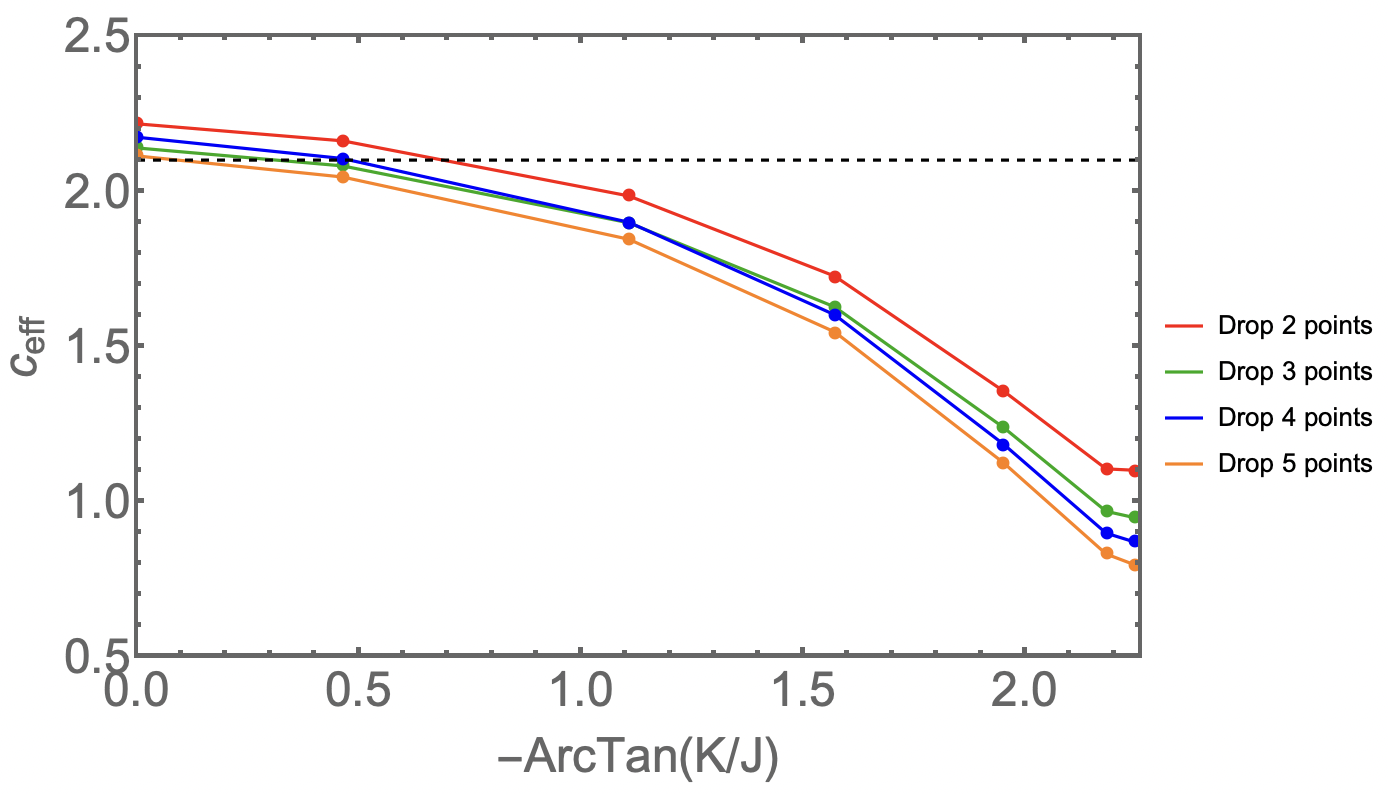}
\caption{The effective central charge of the three chain model in the $K/J<0$ region. This is a gapped phase. 
The results are obtained using $L=20$ and maximum bond dimension $\chi=1500$.}
\label{fig:3chaingap}
\end{figure}

We now study the $N=3$ model numerically. We first study the effective central charge as a function of $K/J$. The result for the $K/J>0$ and $K/J<0$ regions are shown in Fig.~\ref{fig:3chaingapless} and Fig.~\ref{fig:3chaingap} respectively.
Since the $N=3$ model has a large local Hilbert space and the corresponding CFTs have large central charges, 
we can only study the model at fairly small system size $L\lesssim 24$ which is too small for a reliable extraction of scaling dimensions from
the finite-size spectrum. 
Therefore, as before, we calculate the effective central charge by peforming a linear fit of the entanglement entropy as a function of $\log(l_{\rm conformal})$.
In practice, we need to drop the points where $l$ is too small, and the final results depend on how many data points were dropped, as indicated by the different colors in the figures. 

In the $K/J>0$ region, we observe a stable conformal phase. We claim that this corresponds to the irrational CFT discussed in the previous section. 
Remarkably, as predicted by conformal perturbation theory, the central charge of the IR irrational CFT is very close to the UV CFT. 
There exists a 2nd order transition from the conformal phase to a gapped phase at $K/J\approx 0.6$. A natural conjecture for this phase transition is that it is described by the CFT consisting of three copies of the tricritical Ising model ($c=2.1$).
In the $K/J<0$ region, the effective central decreases sharply, indicating that this is a gapped phase.

\begin{figure}[ht]
\centering
\includegraphics[scale=0.4]{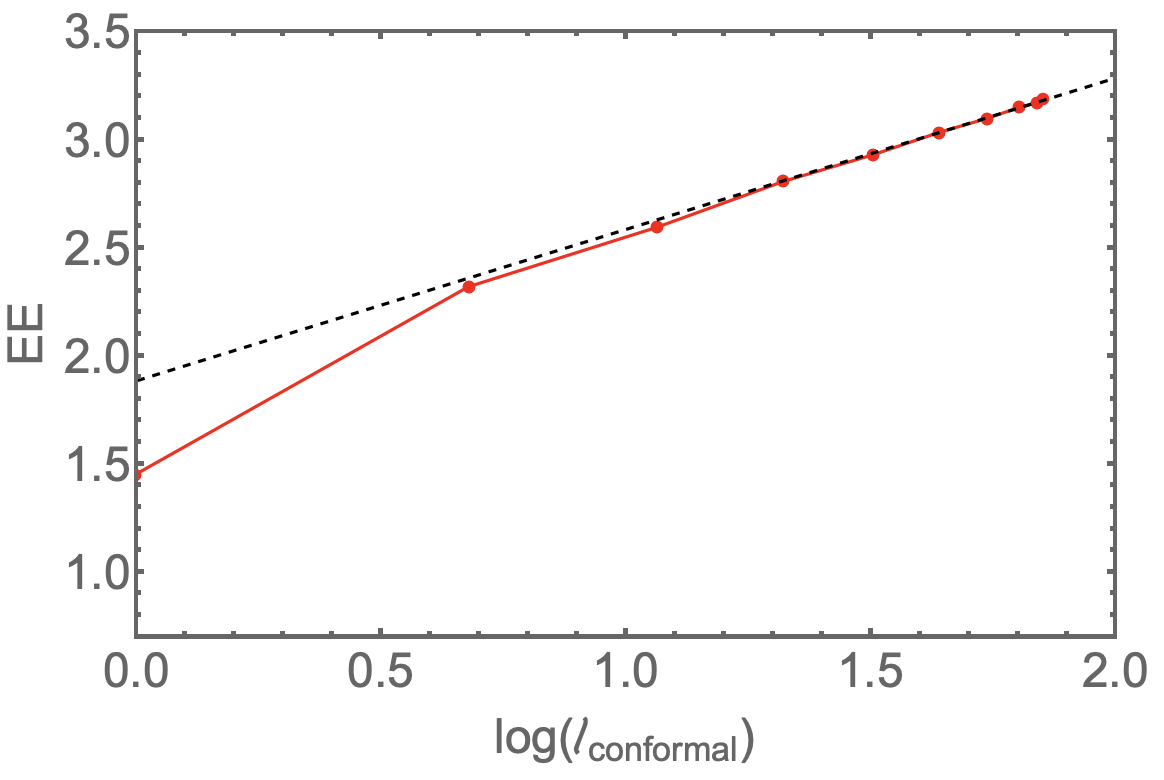}
\caption{The entanglement entropy measure at $K/J=0.25$. 
The results are calculated using $L=24$ and maximum bond dimension $\chi=2400$.}
\label{fig:EE3chain}
\end{figure}

We can try to further increase the lattice size to $L=24$, and use higher bond dimension ($\chi=2400$) to study the irrational CFT more carefully. We set $K/J=0.25$, and the result is given in Fig.~\ref{fig:EE3chain}.
From the entanglement entropy, we get the central charge
\begin{equation}
    c=2.10\pm 0.03.
\end{equation}
This is calculated by performing a linear fit after dropping the five points with the smallest value of $l$. 
Due to the small lattice size, we are not yet able to perform a more sophisticated finite-size analysis; we will leave this for future work. Strictly speaking, our finite-size numerical simulation alone cannot rigorously exclude an extremely weak first-order scenario. Notice, however, the central charge calculated above is very close to the perturbative result in \eqref{perturbativecentral}. Together, the numerical simulation and the perturbative calculation provide strong evidence that the IR phase is indeed a CFT.

\begin{figure}[h]
\centering
\includegraphics[scale=0.45]{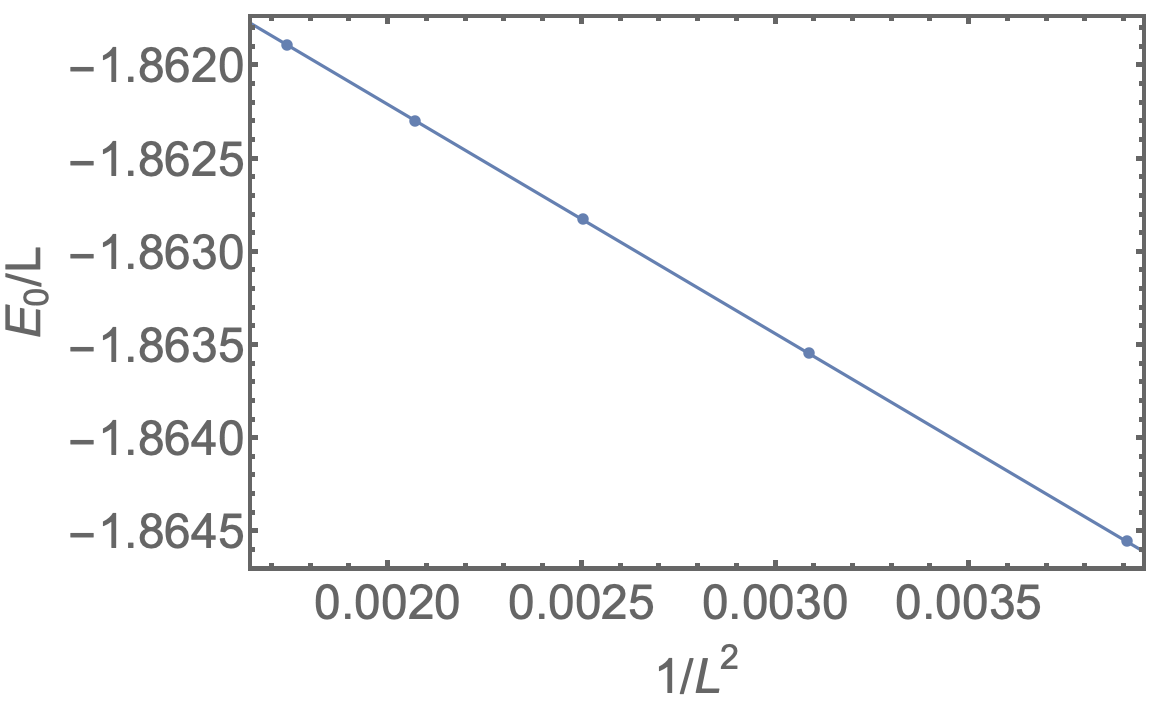}
\caption{The ground state energy density $E_0/L$ versus the (squared) lattice size $L^2$ at the irrational conformal phase.
The data is calculated at $K/J=0.25$.}
\label{fig:Egl3chain}
\end{figure}

Similarly to the $N=2$ chains, we can calculate the dynamical critical exponent of the conformal phase by checking how the ground state energy depends on the lattice size $L$. The result is shown in Fig.~\ref{fig:Egl3chain}.
This shows that the critical phase is indeed a Lorentzian symmetric conformal field theory.

\subsection*{Comments on the Landau-Ginzburg formulation}
Let us briefly comment on a possible Landau-Ginzburg realization of the same CFT.
As is well known, the diagonal Virasoro minimal models $\mathcal{M}_{m,m+1}$ can also be realized as the IR fixed point of a non-compact scalar with a multi-critical potential \cite{Zamolodchikov:1987ti}
\begin{equation}
    \mathcal{L}_{m}=\int d^2x \,\frac{1}{2}(\partial\phi)^2 + \lambda :\phi^{2(m-1)}:\,,
\end{equation}
where several operator identifications can be made, for example the relevant operators $\phi \equiv\phi_{(2,2)}$, $\phi^{2(m-2)}\equiv\phi_{{(1,3)}}$ and the irrelevant operator $\phi^{2(m-1)}\equiv\phi_{{(3,1)}}$.

It is therefore natural to consider the coupled action \eqref{actionfield} and use the Landau-Ginzburg mapping to identify the Lagrangian
\begin{equation}
    \mathcal{L}_{N-\rm{coupled}}= \int d^2x \sum_{i=1}^N \left( \frac{1}{2}(\partial\phi_i)^2 + \lambda_1 :\phi_i^6: \right) + \lambda_2\sum_{i<j}^N :\phi_i^3\phi_j^3:\,,
\end{equation}
 which can be studied in the $3-\epsilon$ expansion \cite{Osborn:2017ucf,Osborn:2020cnf}.
 It is natural to conjecture that there exists a fixed point for this Lagrangian that can be dimensionally continued to our CFT in 2 dimensions.

\section{Comparison to known RCFTs}
\label{sec:list}
We now attempt to list candidate RCFTs that could describe fixed points in our phase diagram. While no rigorous classification of RCFTs in increasing order of $c>1$ exists,\footnote{An orthogonal classification scheme, where one instead increases the number of chiral primaries has produced interesting results \cite{Mathur:1988na,Mukhi:2022bte}, but is not sufficient for our purposes.} all known RCFTs can be built as cosets of WZW models or obtained from Narain moduli spaces \cite{Yin:2017yyn}. We will use two major classes of constructions to generate infinite families of RCFTs. These are cosets of the type
\begin{equation}
    \frac{g_k}{U(1)^{r_g}}\,, \qquad \frac{g_k \oplus g_l}{g_{k+l}}\,,
\end{equation}
where $r_g$ denotes the rank of $g$.\footnote{This list is not claimed to be exhaustive in general, but we believe it is exhaustive in the range $c\leq1.2$. For example, more general cosets of the type $G_k/H_k$ with $H$ a subgroup of $G$ such as $G=SU(N)$ and $H=SU(M)$ with $M<N$ also exist but have quickly growing central charges. 
There are also exceptional conformal embeddings (the so-called Maverick cosets \cite{Dunbar:1993hr}) which are believed to exist only for finitely values of the rank, but all known cases have central charges coinciding with the two main families discussed above.} 
We begin by listing the RCFTs in the range $1<c\leq1.2$. In this interval the list contains only three values of $c$:
\begin{itemize}
    \item $c=8/7\approx1.14$ corresponding to the $\mathbb{Z}_5$ parafermion and its bosonized coset $SU(2)_5/U(1)$ \cite{Fateev:1985mm,Dotsenko:2002gs}.
    \item $c=81/70\approx1.15$ corresponding to the third $\mathcal{N}=1$ super-Virasoro minimal model and its bosonized coset $SU(2)_2\times SU(2)_3/SU(2)_5$ \cite{Goddard:1986ee}.
    \item $c=6/5=1.2$ corresponding to the second $\mathcal{W}_3$ minimal model \cite{Zamolodchikov:1985wn} ($SU(3)_1\times SU(3)_2/SU(3)_3$), as well as the coset $SU(3)_2/U(1)^2$ and the tensor product theory $\textrm{Ising} \otimes \textrm{Tricritical Ising}$.
\end{itemize}
Unfortunately, the sparseness of theories found for $c\leq1.2$ quickly subsides as we increase the central charge. In fact, we find several accumulation points from below in $c$. Below, we list these infinite families of theories along with some finite number of exceptional cases. We should always keep in mind possible additional orbifolds of these theories and/or their non-diagonal modular invariants.
\sloppy
\begin{itemize}
    \item Accumulation at $c\to3/2=1.5$ from $\mathcal{N}=1$ super-Virasoro minimal models with $c=\frac{3}{2}(1- \frac{8}{(k+2)(k+4)})$, with the same central charge as the GSO projected bosonic coset $SU(2)_2\times SU(2)_k/SU(2)_{k+2}$. The even subset of these has the same central charge as the series $SO(3)_1\times SO(3)_k/SO(3)_{k+1}$. At the accumulation we have the pure WZW model $SU(2)_2$.
    \item Accumulation at $c \to 9/5=1.8$ from $SU(2)_3\times SU(2)_k/SU(2)_{k+3}$ cosets with the central charges $c=\frac{9}{5}(1-\frac{10}{(k+2)(k+5)})$. At the accumulation we have the pure WZW model $SU(2)_3$.
    \item Multiple accumulation families as $c\to2$: $\mathcal{W}_3$ minimal models ($SU(3)_1\times SU(3)_k/SU(3)_{k+1}$) with $c=2(1-\frac{12}{(k+3)(k+4)})$; $SU(2)_4\times SU(2)_k/SU(2)_{k+4}$ cosets with $c=2(1-\frac{12}{(k+2)(k+6)})$; $\mathbb{Z}_k$ parafermions ($SU(2)_k/U(1)$) with $c=2(1-\frac{3}{k+2})$; Type II parafermions \cite{Dotsenko:2003ui,Dotsenko:2003zc} ($SO(3)_2\times SO(3)_k/SO(3)_{k+2}$) with  $c=2(1-\frac{3}{(k+1)(k+3)})$; Infinite number of CFTs at $c=2$: Narain moduli space and orbifolds thereof. 
    \item A finite number of $SU(2)_5\times SU(2)_k/SU(2)_{k+5}$ cosets with $5 \leq k \leq 22$ with central charge $c=\frac{15}{7}(1-\frac{14}{(2+k)(7+k)})$. A finite number of $SU(2)_6\times SU(2)_k/SU(2)_{k+6}$ cosets with $6 \leq k \leq 10$ with $c=\frac{9}{4}(1-\frac{16}{(2+k)(8+k)})$ as well as $SU(2)_7\times SU(2)_k/SU(2)_{k+7}$ cosets with $7 \leq k \leq 8$ and $c=\frac{7}{3}(1-\frac{18}{(2+k)(9+k)})$. 
    \item $SU(3)_2\times SU(3)_2/SU(3)_{4}$ with $c=64/35 \approx1.8286$ but has same $c$ as previously mentioned coset. $SU(4)_1\times SU(4)_k/SU(4)_{k+1}$ with $k=1,2,3$ where once again the values of $c$ have appeared before. The same holds for $SU(5)_1\times SU(5)_k/SU(5)_{k+1}$ with $k=1,2$ and $SU(6)_1\times SU(6)_1/SU(6)_{2}$ but  $SU(6)_1\times SU(6)_2/SU(6)_{3}$ with $c=25/12\approx2.0833$ is new. All remaining $SU(N)_1\times SU(N)_1/SU(N)_{2}$ have $c<2$ but with same value as parafermions \cite{Bais:1987zk}. 
    \item  $SO(3)_3\times SO(3)_k/SO(3)_{k+3}$ with $k=3,4,5$. Only $k=4$ is new with $c=81/40=2.025$. $SO(5)_1\times SO(5)_k/SO(5)_{k+1}$ with $k=1,2,3,4,5$ but no new value of $c$. $SO(7)_1\times SO(7)_k/SO(7)_{k+1}$  $k=1,2,3$ with $c=49/24\approx2.0417$ for $k=3$ being new. All $sp(2n)$ cosets reduce to previous values of $c$. $SO(2n)$ cosets also reduce to previous values of central charge.
    \item Cosets of the type $(E_6)_1\times(E_6)_k/(E_6)_{k+1}$ and  $(E_7)_1\times(E_7)_k/(E_7)_{k+1}$ with $k=1,2,3$ are also in range but have no new values of $c$. $(E_8)_1\times(E_8)_k/(E_8)_{k+1}$ with $k=3,4,5$ have $c=256/187\approx 1.3690\,,\,c=208/119 \approx  1.7479\,,\, c=44/21\approx 2.0952$ and are all new. $(E_8)_2\times(E_8)_2/(E_8)_{4}$ has a value of $c$ which appeared previously. $(F_4)_1\times(F_4)_k/(F_4)_{k+1}$ with $k=1$ has same $c$ as above but $k=2$ with $c=91/55\approx1.6546$ is new. $(G_2)_1\times(G_2)_k/(G_2)_{k+1}$ with $k=1,2,3$ have $c$ which already appeared but $k=4$ with $c=91/45\approx2.0222$ is new.
\end{itemize}
\fussy
We will now argue that none of these models can realize a 
$( (\rm{Fibonacci})^3 \rtimes S_3)$
symmetry in a straightforward manner (i.e. via Verlinde lines). In a coset model, the fusion rules are determined by the fusion rules of the primaries in the constituent WZW models. Using this, it is easy to convince one-self that the only way to obtain a line with a Fibonacci fusion rule is to already have such a line at the level of the individual factors. To further endow these lines with an $S_N$ permutation structure, we need to be able to swap the factors in the coset, but this is only possible for the two factors in the numerator, and only when they are equal. We therefore conclude that we can only straightforwardly realize  $( (\rm{Fibonacci})^N \rtimes S_N)$ symmetry via cosets when $N=2$. We will further comment on this in Section \ref{sec:fib2} below.

\medskip

An obvious absence from the list above are tensor product of minimal models with $c=2-\frac{6}{n(n+1)}- \frac{6}{m(m+1)}$ some of which can coincide with central charge values of previously listed cosets but most of which don't (there are accumulation points for every $c=1+6/(m(m+1))$ and even accumulation points of accumulation points at $c=2$!\footnote{See \cite{Belin:2025qjm} for an interesting perspective on this point.}). Such tensor product theories have multiple stress tensors. 
More generally, cosets of the form
\begin{equation}
    \frac{g_k\oplus g_l\oplus g_m}{g_{k+l+m}}\,,
\end{equation}
and its multi-copy generalizations, where $( (\rm{Fibonacci})^N \rtimes S_N)$ topological lines could naturally be realized,
always have $N$ stress-tensors. In our case however, if our CFT were a tensor product of three identical minimal models, the central charge of that theory would be below $c=0.7$ and above $c=0.5$, and such theories do not exist. 

The upshot from this analysis is that none of the rational theories listed above realize our non-invertible symmetry in any obvious way. It is therefore reasonable to conjecture that the $N=3$ model flows to an irrational CFT.
\subsection{CFTs with $(\rm{Fibonacci})^2 \rtimes \mathbb{Z}_2$ symmetry }
\label{sec:fib2}
Our $N=2$ models preserve a $(\rm{Fibonacci})^2 \rtimes \mathbb{Z}_2$ symmetry, and as we argued above, it is straightforward to realize such a symmetry in a coset CFT
\begin{equation}
\label{cosetsq}
    \frac{G_k\times G_k}{G_{2k}}\,,
\end{equation}
if the WZW model $G_k$ already contains a Fibonacci Verlinde line $W$ associated to a primary in a representation $\rho_k$ and if $(\rho_k,0,0)$ is allowed in the coset. We are aware of four WZW models where such $W$ lines are present
\begin{equation}
    G_k= SU(2)_3\,, \quad SU(3)_2\,,\quad (G_2)_1\,, \quad (F_4)_1\,,
\end{equation}
respectively for the primaries in the representations
\begin{equation}
    \rho_k= (\mathbf{3})_3\,, \quad(\mathbf{8})_2\,, \quad (\mathbf{7})_1\,,\quad (\mathbf{26})_1\,,
\end{equation}
where the boldface numbers denote the dimension of the representation. When we consider the coset \eqref{cosetsq} for the last two groups we get central charges corresponding to Virasoro minimal models with $c<1$. Instead, for $SU(2)_3$ we get the coset with $c=1.35$ identified with CFT$_2$, which has also has an additional diagonal $\mathbb{Z}_2$ symmetry. Finally, for $SU(3)_2$ we get $c\approx1.83$ as well as a diagonal $S_3$ symmetry, which agrees with the symmetries of our unknown CFT$_1$. Unfortunately, the value of the central charge seems to be strictly outside our error bars for $c_1\approx1.77$.
\section{Future directions}
\label{sec:Future}
In this work, we have conjectured the existence of compact irrational CFTs in the phase diagram of $N=3$ coupled tricritical Ising models by studying coupled anyonic chains with DMRG. 
Some indirect evidence for irrationality was given using the knowledge of RCFTs with a similar central charge, along with their symmetries, to essentially rule out all the known candidate theories. Of course, there is no rigorous classification of RCFTs with $c>1$ and a more robust method is needed. 
As discussed previously, such RCFTs must have conserved currents other than the stress-tensor, and therefore performing a DMRG analysis at fixed momentum is key for excluding the existence of currents with low spin.\footnote{More generally, understanding the space of consistent chiral algebras, and in particular how large the spin of the leading non-trivial current can be would be a very useful benchmark in this process.}
A particularly useful algorithm is the periodic uniform MPS method~\cite{Zou:2017zce,PhysRevB.83.125104}, which allows one to study states with periodic boundary conditions and fixed momentum.

In addition to $N$ copies of the tricritical Ising model coupled through the $\sigma'$ operator \eqref{actionfield}, we have also studied the $N$ copies of the 3-state Potts models coupled through the $Z$ operator \eqref{actionfield2}. 
In the large $N$ limit, such deformations can be viewed as double-trace deformations of the decoupled theory~\cite{Giombi:2018vtc}. 
Depending on whether the double-trace deformation is relevant or irrelevant, it leads to renormalization group flow towards a CFT in UV or IR.
Classic examples are the flow between free scalar theory and the critical O(N) CFT, and the flow between free fermion theory and the critical Gross-Neveu model.
We expect these corresponding CFTs of \eqref{actionfield} and \eqref{actionfield2} to exist at least in the large $N$ limit.
We have also shown here that the CFTs exist at $N=3$ for the tricritical Ising model and $N=2$ for the coupled Potts model. 
This leads us to conjecture that the CFT exists for any larger $N$, giving us two families of irrational CFT candidates.

A closely related system to consider, is the 3-state Potts models coupled by the energy operator, whose classical lattice version was studied in \cite{Dotsenko:1998gyp}. While the standard local Hilbert space in which this system is formulated is 3-dimensional, which should further complicate our numerics, it is possible to flow to the same fixed point in a locally 2-dimensional Hilbert space \cite{Jacobsen:2024jel}, by making use of the Temperley-Lieb algebra. 
Redoing our analysis in that setup seems promising, since other methods have been used in the past and are available for comparison, namely the $(Q-2)$-state expansion at fixed $N$, which makes it a perturbatively controlled fixed point. In fact, treating $Q$ as a continuous parameter allows us to study a family of fixed points containing both the examples considered in this paper and in \cite{Dotsenko:1998gyp}, as well as potentially infinitely more examples corresponding to coupled minimal models of higher central \cite{AJR}.

In this paper, we focus on a lattice model with $\mathcal{C}$-invariant coupling. 
Considering a more general lattice model with an additional $\mathcal{C}$-invariant coupling would allow us to tune away the leading irrelevant corrections to the fixed points, improving the numerical accuracy, and is therefore a worthy endeavor. This trick has proven to be effective in both Monte Carlo simulations~\cite{PhysRevB.65.144520,PhysRevB.63.214503} and the recent fuzzy sphere regularization approach to study CFTs~\cite{Zhu:2022gjc}.

Finally, we should mention other non-perturbative methods that are available to study our system, and that could complement our analysis. 
It will be interesting to explore the possibility of studying our coupled anyons Hamiltonian using quantum Monte Carlo simulations.
This approach has the advantage that it would allow us to reach much larger system sizes, and might even allows us to consider the generalization of our model to $N\geq4$ coupled chains. Orthogonally, we can also try to study the CFT in the IR directly, by making use of bootstrap methods. 
In two dimensions, two versions of the bootstrap could be useful: The sphere four-point bootstrap,\footnote{See \cite{Kousvos:2024dlz} for a bootstrap study of the coupled Potts CFT.} where it is possible to impose the invertible global symmetry as well as some gaps motivated by our DMRG study; and the modular bootstrap, which studies the partition function on the torus,\footnote{We do not need to assume a finite modular tensor category for these candidate irrational CFTs. Modular invariance of the full torus partition function is expected on general grounds for a local, (non-anomalous) two-dimensional CFT; understanding the corresponding non-rational modular data is an interesting open direction.} and where non-invertible symmetries can be implemented directly \cite{Lin:2023uvm}.\footnote{See also \cite{Nakayama:2025mrm} for a first four-point bootstrap study using conditions motivated by non-invertible symmetry.} We are currently actively investigating this direction.

\paragraph{Acknowledgements}
We thank Connor Behan, Nathan Benjamin, Eduardo Castro, Luke Corcoran, Cl\'ement Delcamp, Paul Fendley, Jesper Jacobsen, Igor Klebanov, Shota Komatsu, Sylvain Lacroix, Jo\~ao Lopes dos Santos, John McGreevy, V\'itor Pereira, Francesco Russo, Slava Rychkov, Hubert Saleur, Julian Sonner, Balt van Rees and Yifan Wang for useful discussions.  AA further thanks Connor Behan for collaboration on related topics and Andreia Gon\c{c}alves for continued support. 
This work was granted access to the HPC resources of IDCS support unit from Ecole polytechnique.
AA is funded by the European Union (ERC, FUNBOOTS, project number 101043588) and was further funded by the project 2024.00230.CERN at the University of Porto.
JR is funded by the European Union (ERC “QFTinAdS”, project number 101087025). 
Views and opinions expressed are however those of the author(s) only and do not necessarily reflect those of the European Union or the European Research Council Executive Agency. Neither the European Union nor the granting authority can be held responsible for them.

\appendix

\section{Review of $SU(2)$ coset CFTs}
\label{app:coset}
In this appendix, we give a brief overview of the properties of the $SU(2)$ coset CFTs
\begin{equation}
    \frac{SU(2)_k\times SU(2)_m}{SU(2)_{k+m}}\,.
\end{equation}
The coset primaries are labeled by triplets of $SU(2)$ representations $(\ell_k,\ell_m,\ell_{k+m})$, where $\ell_p \in \{0,1/2,\dots,p/2\}$, subject to the identification $(\ell_k,\ell_m,\ell_{k+m})\equiv (k/2-\ell_k,m/2-\ell_m,(k+m)/2-\ell_{k+m})$. Furthermore, not all triplets correspond to allowed representations, and are further subject to the constraint
\begin{equation}
    \ell_k+\ell_m-\ell_{k+m}\in \mathbb{Z} \,.
\end{equation}
As is well known, the stress-tensor of the coset model is given by the difference of the numerator and denominator theories' stress tensor (i.e. $T_{\textrm{coset}}=T_k+T_m-T_{k+m}$), and therefore the chiral dimensions of coset primaries satisfy 
\begin{equation}
    h_{(\ell_k,\ell_m,\ell_{k+m})}= h_{\ell_k}+h_{\ell_m}-h_{\ell_{k+m}}+n\,,  n \in \mathbb{N}\,,
\end{equation}
where the addition of a positive integer might be needed since the primary does not always appear in the ground state of the denominator representation. Here, $h_{\ell_k}$ are the weights of primaries in the $SU(2)_k$ WZW model \cite{WESS197195,Witten:1983tw} and are given by
\begin{equation}
    h_{\ell_k}= \frac{\ell_k(\ell_k+1)}{k+2}\,.
\end{equation}
Given the finite list of chiral primaries, we can try to build different modular invariants for the coset theory. For example, if we observe sets of chiral dimensions which differ by an integer, a non-diagonal modular invariant might exist. A large set of modular invariants can be obtained by using the $ADE$-classified modular invariants of each constituent $SU(2)$ factor. Let us consider for example the coset
\begin{equation}
\label{coset235}
    \frac{SU(2)_2\times SU(2)_3}{SU(2)_{5}}\,,
\end{equation}
which has $c\approx1.15$. Since for $k=2$ and $k$ odd, there is only an $A$-type modular invariant for $SU(2)_k$, we see that the coset does not admit non-diagonal modular invariants. Therefore, we can only have scalar primaries with $\Delta_{(\ell_2,\ell_3,\ell_5)}= 2 h_{(\ell_2,\ell_3,\ell_5)}$, i.e. a diagonal modular invariant.
In fact, it is easy to see no integer differences between weights occur,\footnote{Instead, one does find half-integer differences, and in particular a chiral dimension $3/2$. This is because the chiral theory admits a fermionic partition function with spin $1/2$ primaries and a super-current. This is nothing but the third $\mathcal{N}=1$ super-Virasoro minimal model.} as can be seen by listing the first few operators as we did in Table \ref{tab:dimlist}.
\begin{table}[h]
\centering

\caption{Leading scaling dimensions in the coset model \eqref{coset235}.}
\begin{tabular}{ll}
\toprule
$(\ell_3,\ell_2,\ell_5)$ & $\Delta_{(\ell_3,\ell_2,\ell_5)}$ \\
\midrule
$(0,0,0)$ & $0$ \\
 $(1/2,0,1/2)$& $\frac{3}{35}\approx0.0857$ \\
 $(1/2,1/2,1)$& $\frac{29}{280} \approx0.1036$ \\
 $(0,1/2,1/2)$& $\frac{9}{56}\approx0.1607$ \\
 $(1,0,1)$&$\frac{8}{35}\approx0.2286$ \\
 $(0,1,1)$& $\frac{3}{7}\approx0.4286$ \\
 $(1/2,1/2,0)$& $\frac{27}{40}=0.675$ \\
 $(1,1/2,1/2)$& $\frac{269}{280}\approx0.9607$ \\
$(1/2,1,1/2)$ & $\frac{38}{35}\approx 1.0857$ \\
\bottomrule
\end{tabular}
\label{tab:dimlist}
\end{table}
 \medskip
 
 Finally, we describe how to fix the fusion rules. Each member of the triplet describing the primary satisfies its respective $SU(2)_k$ fusion rule, which is given by
\begin{equation}
    \ell_k\times\ell_k'=\sum_{\ell_m=|\ell_k-\ell_k'|}^{\textrm{min}(\ell_k+\ell_k',\, k-\ell_k-\ell_k')} \ell_m \,.
\end{equation}
As a representative example, we see that for $SU(2)_3$ the primary associated to the spin-1 representation of $SU(2)$ satisfies
\begin{equation}
    (1)_3\times(1)_3=(0)_3+(1)_3\,,
\end{equation}
which is Fibonacci-like and will lead to actual Fibonacci topological lines in the case of the $SU(2)_3\times SU(2)_3/SU(2)_6$ coset explained below.
\subsection{RG instability of $SU(2)_3\times SU(2)_3/SU(2)_6$ coset CFT}
\label{app:RG}
We will now show that the $SU(2)_3\times SU(2)_3/SU(2)_6$ coset, which has categorical symmetry $\mathcal{C}$, has a relevant singlet operator under $\mathcal{C}$. First, we need to identify how the symmetries in this model act. The complex conjugation $\mathbb{Z}_2$ acts on primary operators through
\begin{equation}
    \phi_{(\ell_3,\ell_3',\ell_6)} \to \phi_{(\ell_3',\ell_3,\ell_6)}\,.
\end{equation}
The remaining symmetries are generated by Verlinde lines $\mathcal{L}_{(\ell_3,\ell_3',\ell_6)}$ which are in one-to-one correspondence with coset primaries $ \phi_{(\ell_3,\ell_3',\ell_6)}$. The diagonal spin-flip $\mathbb{Z}_2$ line $\eta$ is generated by
\begin{equation}
\mathcal{L}_{(3/2,3/2,0)}\equiv\eta\,, \qquad\eta^2=1\,.
\end{equation}
Similarly, the two Fibonacci symmetries satisfy
\begin{equation}
    \mathcal{L}_{(1,0,0)}\equiv W^{(1)}\,,\quad\mathcal{L}_{(0,1,0)}\equiv W^{(2)}\,,\qquad (W^{(i)})^2=1+W^{(i)}\,,
\end{equation}
and are exchanged by the action of the complex conjugation $\mathbb{Z}_2$. The action of such Verlinde line on coset primaries is well-known to be given by \cite{Verlinde:1988sn}
\begin{equation}
    \mathcal{L}_{(m_3,m_3',m_6)} |\phi_{(\ell_3,\ell_3',\ell_6)}\rangle = \frac{\mathcal{S}_{(m_3,m_3',m_6)}^{(\ell_3,\ell_3',\ell_6)}}{\mathcal{S}_{(0,0,0)}^{(\ell_3,\ell_3',\ell_6)}}|\phi_{(\ell_3,\ell_3',\ell_6)}\rangle\,,
\end{equation}
where $\mathcal{S}_{(m_3,m_3',m_6)}^{(\ell_3,\ell_3',\ell_6)}$ is the modular S-matrix of the coset theory, which is fixed in terms of the constituent $SU(2)$ WZW S-matrices through
\begin{equation}
\mathcal{S}_{(m_3,m_3',m_6)}^{(\ell_3,\ell_3',\ell_6)} = \mathcal{S}_{m_3}^{\ell_3}\mathcal{S}_{m_3'}^{\ell_3'}( \mathcal{S}_{m_6}^{\ell_6})^{-1}\,,
\end{equation}
with the $SU(2)_k$ S-matrices
\begin{equation}
    \mathcal{S}_\ell^{\ell'}= \sqrt{\frac{2}{k+2}} \sin \left( \pi\frac{(2\ell+1)(2\ell'+1)}{k+2}\right)\,.
\end{equation}
For a local operator $\phi$ to be invariant under the action of a Verlinde line $\mathcal{L}$, it must acted upon in the same way as the identity operator. Therefore, the invariance condition reads
\begin{equation}
    \frac{\mathcal{S}_{(m_3,m_3',m_6)}^{(\ell_3,\ell_3',\ell_6)}}{\mathcal{S}_{(0,0,0)}^{(\ell_3,\ell_3',\ell_6)}}=\frac{\mathcal{S}_{(m_3,m_3',m_6)}^{(0,0,0)}}{\mathcal{S}_{(0,0,0)}^{(0,0,0)}}\,.
\end{equation}
\fussy
It is then easy to check that the operator $\phi_{(3/2,3/2,2)}$ is invariant under the three Verlinde lines, and obviously over the conjugation symmetry. It satisfies \begin{equation}
    h_{(3/2,3/2,2)}=\bar{h}_{(3/2,3/2,2)}=3/4\,,
\end{equation} and is therefore a relevant singlet, from which we conclude that $SU(2)_3\times SU(2)_3/SU(2)_6$ is not the stable fixed point of our model, but can potentially be reached by tuning an additional $\mathcal{C}$ invariant coupling, i.e. it is the obvious candidate to be the CFT$_2$ in our phase diagram.

\section{Finite size correction to the effective central charge}
\label{app:finitesize}
\begin{figure}[htbp]
    \centering
    \begin{subcaptionbox}{Entanglement entropy\label{fig:a}}[0.43\linewidth]
        {\includegraphics[width=\linewidth]{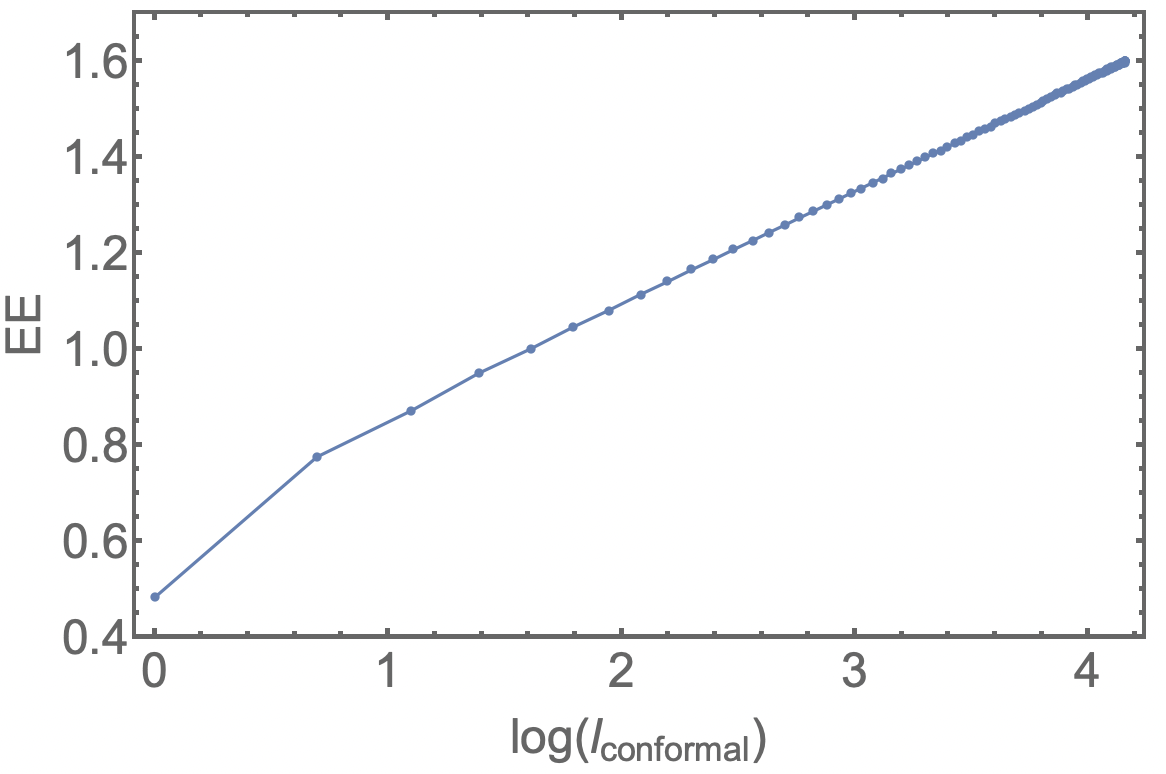}}
    \end{subcaptionbox}
    \hfill
    \begin{subcaptionbox}{Effective central charge\label{fig:b}}[0.45\linewidth]
        {\includegraphics[width=\linewidth]{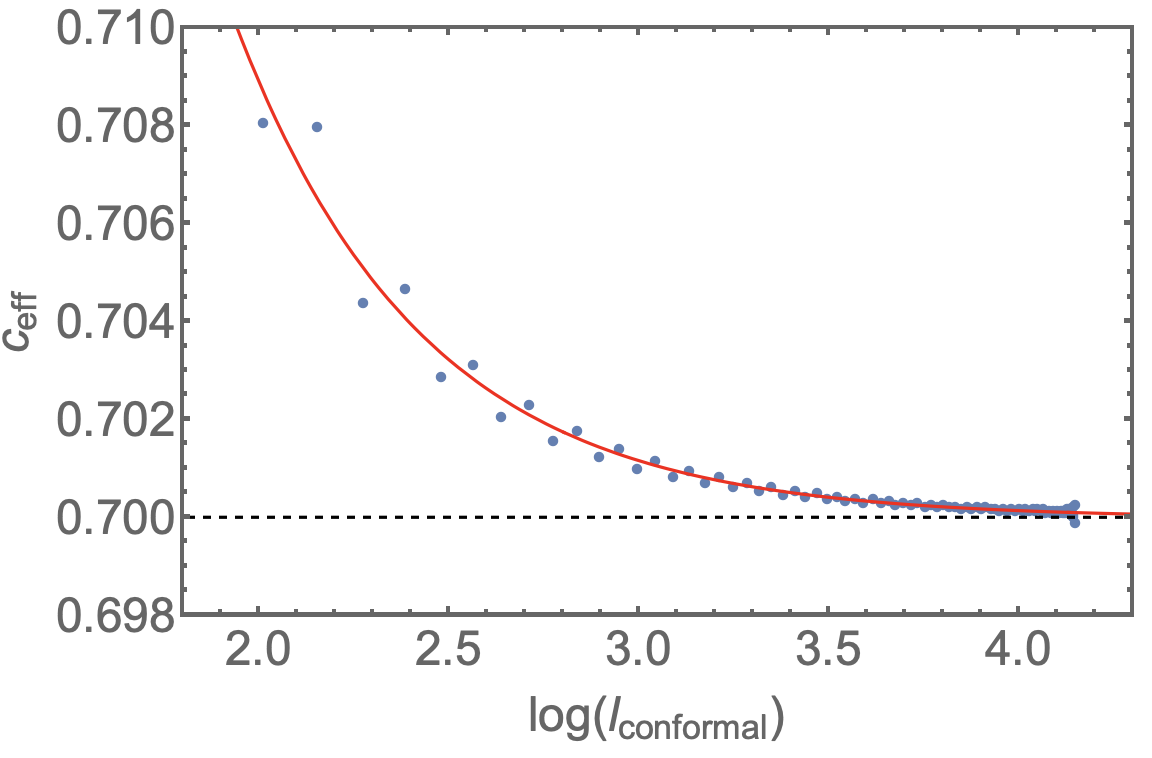}}
    \end{subcaptionbox}
\caption{Left: The Entanglement entropy of a single anyon chain, fitted using $EE(t)= a\cdot e^{-2 t}+ c_{\infty} t + b$, which corresponds to the red line. 
The effective central charge on the right plot is defined as $c_{\rm eff}(t)=\frac{dEE}{dt}$. 
Right: The effective central charge fitted using $c(t)=c_{\infty}+a e^{-2\omega t}$. The data points are obtained by doing a linear fit of eight neighboring points in the left plot and extracting the slope. 
The three points near the boundary are discarded because they are likely to be affected by the UV details of the lattice. We obtain $c_{\infty}=0.69998(4)$ and $\omega=1.01(4)$.}
    \label{fig:side_by_side}
\end{figure}

In this appendix, we discuss how to improve estimates to the effective central charge by performing a finite size scaling analysis.
To estimate these finite size corrections we consider a toy beta function
\begin{align}
    l \frac{d}{dl} g= -\omega g+\mathcal{O}(g^2).
\end{align}
Such a beta function naturally appears in perturbation theory with a weakly relevant operator.
Perturbative RG flow is a gradient flow, and one can define a potential~\cite{Jack:1990eb}
\begin{align}
    A(g)=-\frac{1}{2}\omega g^2+\mathcal{O}(g^2).
\end{align}
This allows us to write $\beta(g)=\frac{d A(g)}{d g}$, as expected from a gradient flow. 
Solving the beta function, we get that 
\begin{align}
    A(t)=2 \omega a\cdot e^{-2\omega t}+a_{\infty},\quad {\rm with}\quad t=\log(l).
\end{align}
Similar to the central charge in 2D CFT, such a potential $A$ is also a measure of the effective degrees of freedom, and in fact is proportional to it in two dimensions \cite{Zamolodchikov:1986gt,Cardy:1988tj}. It is therefore reasonable to assume that the central charge has the same finite size corrections.
The effective central charge is the derivative of the entanglement entropy with respect to $t=\log(l)$. We therefore have 
\begin{align}
    EE(t)= a\cdot e^{-2\omega t}+ c_{\infty} t + b,
\end{align}
where $a, b$ are non-universal constant, while $c_{\infty}$ and $\omega$ depend on the CFT data. The constant $c_{\infty}$ is just the central charge, while $\omega$ is controlled by the leading irrelevant operator $\epsilon'$ (invariant under the Fibonacci symmetry), that is $\omega=D-\Delta_{\epsilon'}$. 
We can test this using a single anyon chain with $J>0$. 
The result in Fig.~\ref{fig:side_by_side} confirms that the entanglement entropy indeed follows the proposed finite-size behavior.

\bibliography{main.bib}
\bibliographystyle{utphys}

\end{document}